\documentclass{aa}  
\usepackage[]{aalongtable}
\usepackage{scalefnt}
\usepackage[usenames]{color}
\usepackage{graphicx}
\usepackage{txfonts}
\def\kms   {km~s$^{-1}$}
\def\loggf {log~$gf$}
\def\Teff  {$T_\mathrm{eff}$}
\def\logg  {log~$g$}
\begin{document}

\title{High-resolution abundance analysis of HD~140283
\thanks{Based on observations within Brazilian time at the 
Canada-France-Hawaii Telescope (CFHT) which is operated by the National 
Research Council of Canada, the Institut National des Sciences de 
l'Univers of the Centre National de la Recherche Scientifique of France, 
and the University of Hawaii; Progr. ID 11AB01.}
\thanks{Tables A.1, A.2, and A.3 are only available in electronic form
at the CDS via anonymous ftp to cdsarc.u-strasbg.fr (130.79.128.5)
or via http://cdsweb.u-strasbg.fr/cgi-bin/qcat?J/A+A/}}


\author{
C.~Siqueira-Mello\inst{1}
\and
S.~M.~Andrievsky\inst{2,3}
\and
B.~Barbuy\inst{1}
\and
M.~Spite\inst{3}
\and
F.~Spite\inst{3}
\and
S.~A.~Korotin\inst{2}
}
\offprints{C. Siqueira Mello Jr. (cesar.mello@usp.br).}

\institute{
Universidade de S\~ao Paulo, IAG, Departamento de Astronomia, 
Rua do Mat\~ao 1226, S\~ao Paulo 05508-900, Brazil
\and
Department of Astronomy and Astronomical Observatory, Odessa National 
University, and Isaac Newton Institute of Chile Odessa branch, 
Shevchenko Park, 65014 Odessa, Ukraine
\and
GEPI, Observatoire de Paris, PSL Research University, CNRS,  
Universit\'e Paris Diderot, Sorbonne Paris Cit\'e, Place Jules Janssen, 
92195 Meudon, France
}
\date{Received 08 June 2015; accepted 10 August 2015}

\titlerunning{}
\abstract
{HD 140283 is a reference subgiant that is metal poor and confirmed to be 
a very old star. The abundances of this type of old star can constrain the nature and 
nucleosynthesis processes that occurred in its (even older) progenitors. 
The present study may shed light on nucleosynthesis processes yielding heavy elements 
early in the Galaxy.}
{A detailed abundance analysis of a high-quality spectrum is carried out, 
with the intent of providing a reference on stellar lines and abundances 
of a very old, metal-poor subgiant. We aim to derive abundances 
from most available and measurable spectral lines.}
{The analysis is carried out using high-resolution (R~$=$~81 000) and high 
signal-to-noise ratio (800~$<$~S/N/pixel~$<$~3400) spectrum, in the wavelength range 
3700~$<\lambda({\AA})<$~10475, obtained with a seven-hour exposure time, using the 
Echelle SpectroPolarimetric Device for the Observation of Stars (ESPaDOnS) at the 
Canada-France-Hawaii Telescope (CFHT). The calculations in local thermodynamic equilibrium (LTE) 
were performed with the OSMARCS 1D atmospheric model and the spectrum synthesis code Turbospectrum, 
while the analysis in non-local thermodynamic equilibrium (NLTE) is based on the MULTI code. 
We present LTE abundances for 26 elements, and NLTE calculations for the species 
\ion{C}{I}, \ion{O}{I}, \ion{Na}{I}, \ion{Mg}{I}, \ion{Al}{I}, \ion{K}{I}, \ion{Ca}{I}, 
\ion{Sr}{II}, and \ion{Ba}{II} lines.}
{The abundance analysis provided an extensive line list suitable for metal-poor subgiant stars. 
The results for Li, CNO, $\alpha$-, and iron peak elements are in good agreement with literature. 
The newly NLTE Ba abundance, along with a NLTE Eu correction and a 3D Ba correction from literature, 
leads to [Eu/Ba]~$=+0.59\pm0.18$. This result confirms a dominant r-process contribution, possibly 
together with a very small contribution from the main s-process, to the neutron-capture 
elements in HD~140283. Overabundances of the lighter heavy elements and the high abundances derived 
for Ba, La, and Ce favour the operation of the weak r-process in HD~140283.}
{}

   \maketitle
%

\section{Introduction}

HD~140283 (V = 7.21; Casagrande et al. 2010) is a subgiant, metal-poor 
star in the solar neighbourhood, which is analysed extensively in the literature, 
and has historical importance in the context of the existence of 
metal-deficient stars (Chamberlain \& Aller 1951). 
In recent decades, element abundances in HD~140283 were studied 
by several authors, and a bibliographic compilation up to 2010 can be found 
in the PASTEL catalogue (Soubiran et al. 2010). Most recently, Frebel \& Norris (2015) 
stressed the importance of this star to the history of the discovery of the most metal-poor 
stars in the halo. More recently, hydrodynamical 
3D models and NLTE computations were applied for lithium lines, for example, by 
Lind et al. (2012) and Steffen et al. (2012). 
The fractions of odd and even barium isotopes in HD~140283 have been the subject of 
intense debate, given that the even-Z isotopes are only produced by the 
neutron capture s-process, whereas the odd-Z isotopes are produced by both 
the s- and r-processes (Gallagher et al. 2010, 2012, 2015). 
Based on UV spectra, the molybdenum abundance in HD~140283 was derived by 
Peterson (2011). Roederer (2012) also used UV lines to obtain abundances of 
zinc (\ion{Zn}{II}), arsenic (\ion{As}{I}), and selenium (\ion{Se}{I}), 
among other elements, in addition to upper limits for germanium (\ion{Ge}{I}) 
and platinum (\ion{Pt}{I}). Siqueira-Mello et al. (2012), hereafter Paper I, 
analysed the origin of heavy elements in HD~140283 deriving the europium 
abundance and making the case for an r-process contribution in this star. 

Bond et al. (2013) derived an age of 14.46$\pm$0.31 Gyr for HD~140283, using 
a trigonometric parallax of 17.15$\pm$0.14 mas measured with the Hubble 
Space Telescope, making this object the oldest known star for which a 
reliable age has been determined. Bond et al. employed evolutionary tracks 
and isochrones computed with the University of Victoria code (VandenBerg et 
al. 2012), with an adopted helium abundance of Y$=$0.250, and including 
effects of diffusion, revised nuclear reaction rates, and enhanced 
oxygen abundance. More recently, VandenBerg et al. (2014) presented a 
revised age of 14.27$\pm$0.38 Gyr. This age is slightly larger than 
the age of the universe of 13.799$\pm$0.038 Gyr  based on the cosmic 
microwave background (CMB) radiation as given by the 
Planck collaboration (Adam et al. 2015). 
According to VandenBerg et al. (2014), uncertainties, particularly in the oxygen 
abundance and model temperature to observed colour relations, can explain 
this difference, but the remote possibility that this object is older than 
14 Gyr cannot be excluded.  HD~140283 is therefore a very old 
star that must have formed soon after the Big Bang. In the future, 
asteroseismology could help to verify the age of HD~140283.

In this work, we carry out a detailed analysis and abundance derivation 
for HD~140283. The main motivation for this study was triggered by the 
controversy discussed above about the interpretation of barium isotopic abundances, 
and the possibility of testing whether heavy elements 
are produced by the r- or s-process. With this purpose, we 
obtained a seven-hour exposure high-S/N spectrum, with a wavelength coverage 
in the range 3700~$<\lambda({\rm \AA})<$~10475 for this star.

In Sect. 2 the observations are reported. 
In Sect. 3 the atmospheric parameters are derived. 
In Sect. 4 the abundances computed in LTE and NLTE are presented. 
In Sect. 5 the results are discussed, and final conclusions are drawn in Sect. 6.

\section {Observations and reductions}

HD~140283 was observed in programme 11AB01 (PI: B. Barbuy) 
at the CFHT telescope with the spectrograph 
ESPaDOnS in Queue Service Observing (QSO) mode, to obtain a spectrum 
in the wavelength range 3700-10475~{\AA} with a resolving power of 
R~$=$~81 000. The observations were carried out in 2011, June 12, 14, 15, 
and 16, and July 8. The total number of 23 individual spectra with 20 min 
exposure each produced a total exposure time of more than seven hours. 
The data reduction was performed using the software 
Libre-ESpRIT, a new release of ESpRIT (Donati et al. 1997), running within 
the CFHT pipeline Upena\footnote{http://www.cfht.hawaii.edu/Instruments/Upena/index.html}. 
This package facilitates reducing all exposures automatically, and further fits  continua 
and normalizes to 1. The co-added spectrum was obtained after radial velocity correction and 
a S/N ratio of 800~$-$~3400 per pixel was obtained. Three spectra were 
discarded because of their lower quality as compared with the average.

\section{Atmospheric parameters}

\subsection{Measurement of equivalent widths}
\label{EW}

To derive the atmospheric parameters and abundances, 
we measured the equivalent widths (EWs) of several iron and titanium 
lines in their neutral and ionized states using a semi-automatic code, 
which traces the continuum and uses a Gaussian profile to fit the 
absorption lines, as described in Siqueira-Mello et al. (2014). The 
code can deal with blends on the wings, excluding the parts of the 
line from the computations. 

\begin{figure}
\centering
\includegraphics[width=\hsize]{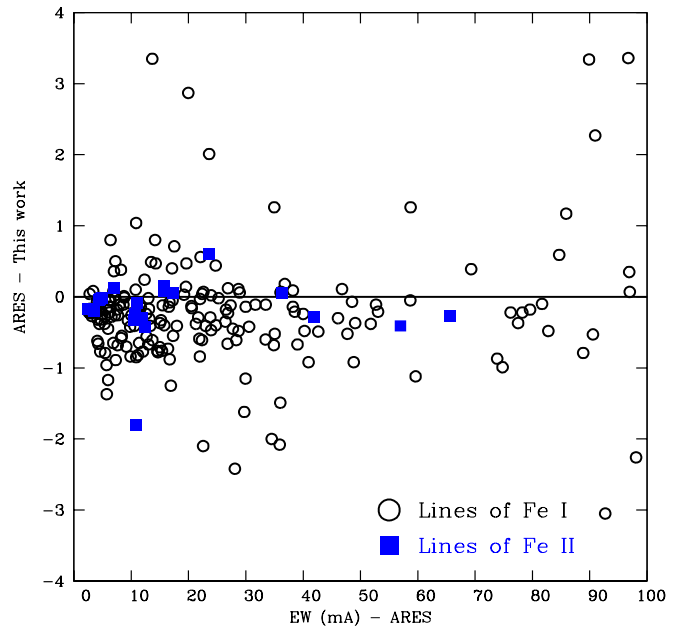}
\caption{Comparison of EWs measured for a set of \ion{Fe}{I} and 
\ion{Fe}{II} lines in HD 140283 using the code by Siqueira-Mello et al.
(2014) and EWs measured using ARES.}
\label{EW_compara}
\end{figure}

To check the reliability of the implemented code, the results were 
compared with those obtained using the routine for the automatic 
measurement of line equivalent widths in stellar spectra ARES 
(Sousa et al. 2007), and only the lines identified by ARES were used to achieve 
the best confidence in the final results. A very good agreement between 
the two measurements is shown in Fig. \ref{EW_compara}. We find a mean 
difference of $\hbox{EW(this work)}-\hbox{EW(ARES)}=-0.22\pm0.43$~m{\AA}, 
which can be considered negligible in terms of EWs.

The complete list of Fe and Ti lines used is shown in Table \ref{EW_measurements}, 
which includes wavelength ({\AA}), excitation potential (eV), \loggf~values from 
the VALD and NIST\footnote{http://physics.nist.gov/PhysRefData/ASD/lines$_-$form.html} 
databases, the measured EWs (m{\AA}), and the derived abundances. Using the errors 
given for the parameters of the Gaussian profile obtained from the fitting 
procedure, the uncertainties $\sigma$EW of the equivalent widths were 
computed based on standard error propagation.

\subsection{Calculations}

Iron and titanium abundances were derived using equivalent widths, 
as usual. All other element abundances  were derived from fits of 
synthetic spectra to the observed spectrum of HD~140283. The OSMARCS 
1D model atmosphere grid  was employed (Gustafsson et al. 2008). 

We used the spectrum synthesis code Turbospectrum (Alvarez \& Plez 1998), 
which includes treatment of scattering in the blue and UV domain, 
molecular dissociative equilibrium, and collisional broadening by H, 
He, and H$_{2}$, following Anstee \& O'Mara (1995), Barklem \& O'Mara (1997), 
and Barklem et al. (1998). The calculations used the Turbospectrum molecular 
line lists (Alvarez \& Plez 1998), and atomic line lists from the VALD 
compilation (Kupka et al. 1999)\footnote{http://vald.astro.univie.ac.at/~vald3/php/vald.php}. 
When available, new experimental oscillator strengths were adopted from 
literature. In addition, hyperfine structure (HFS) splitting and isotope 
shifts were implemented when needed and available.

We also performed NLTE calculations for the following ions: \ion{C}{I}, 
\ion{O}{I}, \ion{Na}{I}, \ion{Mg}{I}, \ion{Al}{I}, \ion{K}{I}, \ion{Ca}{I}, 
\ion{Sr}{II}, and \ion{Ba}{II}. Atomic models for these species were used 
in combination with the NLTE MULTI code (Carlsson 1986; Korotin et al. 1999), 
which facilitates a very good description of the radiation field. The updated 
version of the MULTI code includes opacities from ATLAS9 (Castelli \& Kurucz 2003), 
which modify the intensity distribution in the UV region.

The lines studied in NLTE are often blended 
with lines of other species. Proper comparison of the synthesized and 
observed profiles thus requires a multi-element synthesis. To accomplish this, 
we fold the NLTE (MULTI) calculations into the LTE synthetic spectrum 
code SYNTHV (Tsymbal 1996). With these two codes, we calculate synthetic 
spectra for each region in the vicinity of the line of interest taking 
(in LTE) all the blending lines located in  this region and listed in 
the VALD database into account. 
For the lines of interest (treated in NLTE), the corresponding departure 
coefficients (so-called $b$-factors: $b = n_{\rm i}/n^{*}_{\rm i}$, the 
ratio of NLTE to LTE atomic level populations) obtained with MULTI code are 
the input to SYNTHV code, where they are used in the calculation of the line 
source function, and then for the NLTE line profile. Other possible blending 
lines are treated in LTE.

\subsection{Stellar parameters}

Following Paper I, we adopted the stellar 
parameters \Teff~$=5750\pm100$~K, $\hbox{[Fe/H]}\footnote{
[X/H]~$=$~A(X)$_{star}-$~A(X)$_{\odot}$}=-2.5\pm0.2$ 
and $\xi=1.4\pm0.1$~\kms~from Aoki et al. (2004) and 
\logg~$=3.7\pm0.1$~[g in cgs] from Collet et al. (2009). 
Using the newly measured EWs, we obtained the iron 
abundances A(\ion{Fe}{I})\footnote{We adopted the notation 
A(X) = log~(X) = log~n(X)/n(H)~+~12, with n = number 
density of atoms.}~$=+4.91\pm0.07$ and A(\ion{Fe}{II})~$=+4.96\pm0.07$, 
or $\hbox{[Fe I/H]}=-2.59\pm0.08$ and $\hbox{[Fe II/H]}=-2.54\pm0.08$, 
using the solar abundance A(Fe)$_{\odot}=+7.50\pm0.04$ 
from Asplund et al. (2009). The results are in very good agreement with the 
literature, as in Gallagher et al. (2010), where [Fe/H]$=-2.59\pm0.06$ was obtained. 
For titanium, we found A(\ion{Ti}{I})~$=+2.71\pm0.01$ 
and A(\ion{Ti}{II})~$=+2.69\pm0.03$, or $\hbox{[Ti I/H]}=-2.24\pm0.05$ 
and $\hbox{[Ti II/H]}=-2.26\pm0.06$, using the solar abundance of 
titanium A(Ti)$_{\odot}=+4.95\pm0.05$ from Asplund et al. (2009). 
In Table \ref{summary} we summarize the atmospheric parameters 
adopted, together with the iron and titanium abundances.

\begin{figure}
\centering
\includegraphics[width=\hsize]{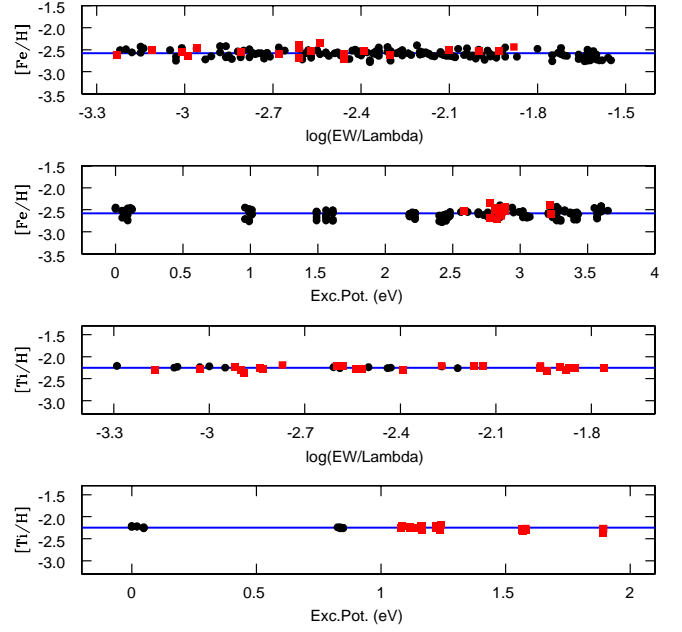}
\caption{Excitation and ionization equilibria of Fe and Ti lines, 
resulting from the set of atmospheric parameters for HD~140283. 
The black dots are the abundances obtained from \ion{Ti}{I} and 
\ion{Fe}{I} lines, the red squares are those from \ion{Ti}{II} 
and \ion{Fe}{II} lines, and the blue solid lines represent the 
average abundances.}
\label{avalia}
\end{figure}

Figure \ref{avalia} shows the dependence of [\ion{Fe}{I}/H], [\ion{Fe}{II}/H], 
[\ion{Ti}{I}/H], and [\ion{Ti}{II}/H] on log~$(EW/\lambda)$, and on the 
excitation potential of the lines obtained for HD~140283. 
The blue solid lines represent the average 
abundances in each case. The excitation and ionization equilibria of 
Fe and Ti lines, resulting from the set of atmospheric parameters adopted 
for HD~140283, confirm the robustness of our choice. However, it should be 
noted that the surface gravity value may be affected by NLTE effects and 
by uncertainties in the oscillator strengths of the Fe and Ti lines.

The broadening parameters in HD~140283 were carefully analysed by 
several authors. We adopted a Gaussian profile 
to take the effects of macroturbulence, rotational, and 
instrumental broadening into account.

\begin{table}
\caption{Atmospheric parameters adopted for HD~140283.}             
\label{summary}      
\scalefont{1.0}
\centering                          
\begin{tabular}{cc}        
\hline\hline                 
\noalign{\smallskip}
\hbox{} & \hbox{Values} \\
\noalign{\smallskip}
\hline
\noalign{\smallskip}
\hbox{\Teff}           & $5750\pm100$~K \\
\hbox{\logg}           & $3.7\pm0.1$~[g in cgs] \\
\hbox{[Fe/H]$_{model}$}& $-2.5\pm0.2$ \\
\hbox{$\xi$}           & $1.4\pm0.1$~\kms \\
\hbox{[\ion{Fe}{I}/H]} & $-2.59\pm0.08$ \\
\hbox{[\ion{Fe}{II}/H]}& $-2.54\pm0.08$ \\
\hbox{[\ion{Ti}{I}/H]} & $-2.24\pm0.05$ \\
\hbox{[\ion{Ti}{II}/H]}& $-2.26\pm0.06$ \\
\noalign{\smallskip}
\hline
\end{tabular}
\end{table}

\subsection{Uncertainties in the derived abundances}

As described in Paper I, the adopted atmospheric parameters present 
typical errors of $\Delta$T$_{\rm eff}=\pm100$~K, $\Delta$\logg~$=\pm0.1$~[g in cgs], 
and $\Delta\xi=\pm0.1$~\kms. Since the stellar parameters are not 
independent from each other, the quadratic sum of the various sources 
of uncertainties is not the best way to estimate the total error budget, 
otherwise it is mandatory to include the covariance terms in this calculation, 
and an estimated correlation matrix may introduce uncontrollable error sources.

To compute the total error budget in the abundance analysis arising 
from the stellar parameters, we created a new atmospheric model with a 100 K 
lower temperature, determining the corresponding surface gravity and 
microturbulent velocity with the traditional spectroscopic method. 
Requiring that the iron abundance derived from \ion{Fe}{I} and \ion{Fe}{II} 
lines be identical, we determined the respective \logg~value, and the microturbulent 
velocity was found requiring that the abundances derived for individual \ion{Fe}{I} 
lines be independent of the equivalent width values. 
The result is a model with T$_{\rm eff}=5650$~K, \logg~$=3.3$~[g in cgs], 
and $\xi=1.2$~\kms. The abundance analysis was carried out with this new model, 
and the difference in comparison with the nominal model should represent the 
uncertainties from the atmospheric parameters. 

Observational errors were estimated using the standard deviation of the abundances 
from the individual lines for each element, and taking into account the uncertainties 
in defining the continuum, fitting the line profiles, and in the oscillator strengths. 
For the elements with only one line available, we adopted the observational error from 
iron as a representative value. The adopted total error budget is the quadratic sum of 
uncertainties arising from atmospheric parameters and observations. 

To estimate the uncertainties in the individual abundances due to the uncertainties 
in the EWs $\sigma$EW, we recomputed the abundances using $\hbox{EW}+\sigma\hbox{EW}$, 
and the differences with respect to the nominal values were adopted as the errors. 
The errors line-by-line are also shown in Table \ref{EW_measurements}.

\begin{table}
\caption{Abundance uncertainties due to stellar parameters $\Delta_{par}$, 
observational errors $\Delta_{obs}$, and adopted total error budget $\Delta_{total}$ 
for LTE abundances.}             
\label{finalabund}      
\scalefont{1.0}
\centering                          
\begin{tabular}{ccrr}        
\hline\hline                 
\noalign{\smallskip}
\hbox{Element} & \hbox{$\Delta_{par}$} & \hbox{$\Delta_{obs}$} & \hbox{$\Delta_{total}$}\\
\noalign{\smallskip}
\hline
\noalign{\smallskip}
\hbox{[Fe/H]}  & $-$0.07 & 0.07 & 0.10 \\
\hbox{[Li/H]}  & $-$0.08 & 0.07 & 0.11 \\
\hbox{[C/Fe]}  & $+$0.01 & 0.07 & 0.07 \\
\hbox{[N/Fe]}  & $-$0.10 & 0.07 & 0.12 \\
\hbox{[O/Fe]}  & $-$0.15 & 0.07 & 0.17 \\
\hbox{[Na/Fe]} & $-$0.04 & 0.08 & 0.09 \\
\hbox{[Mg/Fe]} & $-$0.04 & 0.07 & 0.08 \\
\hbox{[Al/Fe]} & $-$0.01 & 0.07 & 0.07 \\
\hbox{[Si/Fe]} & $-$0.08 & 0.01 & 0.08 \\
\hbox{[K/Fe]}  & $-$0.06 & 0.07 & 0.09 \\
\hbox{[Ca/Fe]} & $-$0.03 & 0.04 & 0.05 \\
\hbox{[Sc/Fe]} & $-$0.07 & 0.06 & 0.09 \\
\hbox{[Ti/Fe]} & $-$0.07 & 0.03 & 0.08 \\
\hbox{[V/Fe]}  & $-$0.07 & 0.08 & 0.11 \\
\hbox{[Cr/Fe]} & $-$0.08 & 0.04 & 0.09 \\
\hbox{[Mn/Fe]} & $-$0.07 & 0.04 & 0.08 \\
\hbox{[Co/Fe]} & $-$0.07 & 0.11 & 0.13 \\
\hbox{[Ni/Fe]} & $-$0.07 & 0.05 & 0.09 \\
\hbox{[Zn/Fe]} & $-$0.07 & 0.07 & 0.10 \\
\hbox{[Sr/Fe]} & $-$0.02 & 0.07 & 0.07 \\
\hbox{[Y/Fe]}  & $-$0.10 & 0.06 & 0.12 \\
\hbox{[Zr/Fe]} & $-$0.11 & 0.05 & 0.12 \\
\hbox{[Ba/Fe]} & $-$0.09 & 0.10 & 0.13 \\
\hbox{[Ce/Fe]} & $-$0.07 & 0.18 & 0.19 \\
\noalign{\smallskip}
\hline
\end{tabular}
\end{table}

\section{Abundance derivation}

Table \ref{linelist} shows the line list of elements other than Fe and 
Ti used in this work, including wavelength ({\AA}), excitation 
potential (eV), adopted oscillator strength, and abundance derived 
from each line. The final LTE abundances derived in HD~140283 for all 
the analysed species are shown in Table \ref{finalabund}. The adopted 
solar abundances from Asplund et al. (2009) are also listed in 
Table \ref{finalabund}. We discuss below each element in terms of 
lines used, HFS splitting, and abundances adopted. In Table 
\ref{fractions} we report the solar isotopic fractions adopted 
from Asplund et al. (2009) that are relevant for HFS computations, 
and in Table \ref{HFSconstants} we summarize the hyperfine coupling 
constants adopted for the lines retained in this analysis.

\begin{table}
\caption{LTE abundances. Column 2 gives the solar abundances from Asplund et al.
(2009), columns 3, 4, and 5 give the absolute abundance
with respect to A(H)=12.0 and the usual logarithmic ratio notation
 with respect to H and to Fe, respectively.}             
\label{finalabund}      
\scalefont{1.0}
\centering                          
\begin{tabular}{ccrrr}        
\hline\hline                 
\noalign{\smallskip}
\hbox{Ion} & \hbox{A(X)$_{\odot}$} & \hbox{A(X)} & \hbox{[X/H]} & \hbox{[X/Fe]} \\
\noalign{\smallskip}
\hline
\noalign{\smallskip}
\hbox{\ion{Fe}{I}}  & 7.50$\pm$0.04 & $+$4.91 & $-$2.59 & ---  \\
\hbox{\ion{Fe}{II}} & 7.50$\pm$0.04 & $+$4.96 & $-$2.54 & ---  \\
\hbox{\ion{Li}{I}}  & 1.05$\pm$0.10 & $+$2.14 & $+$1.09 & ---  \\
\hbox{C(CH)}        & 8.43$\pm$0.05 & $+$6.30 & $-$2.13 & $+$0.46 \\
\hbox{\ion{C}{I}}   & 8.43$\pm$0.05 & $+$6.44 & $-$1.99 & $+$0.60 \\
\hbox{N(CN)}        & 7.83$\pm$0.05 & $+$6.30 & $-$1.53 & $+$1.06 \\
\hbox{[\ion{O}{I}]} & 8.69$\pm$0.05 & $+$6.95 & $-$1.74 & $+$0.85 \\
\hbox{\ion{O}{I}}   & 8.69$\pm$0.05 & $+$7.11 & $-$1.58 & $+$1.00 \\
\hbox{\ion{Na}{I}}  & 6.24$\pm$0.04 & $+$3.62 & $-$2.62 & $-$0.04  \\
\hbox{\ion{Mg}{I}}  & 7.60$\pm$0.04 & $+$5.27 & $-$2.33 & $+$0.26  \\
\hbox{\ion{Mg}{II}} & 7.60$\pm$0.04 & $+$5.66 & $-$1.95 & $+$0.64  \\
\hbox{\ion{Al}{I}}  & 6.45$\pm$0.03 & $+$2.96 & $-$3.50 & $-$0.91 \\
\hbox{\ion{Si}{I}}  & 7.51$\pm$0.03 & $+$5.30 & $-$2.21 & $+$0.38  \\
\hbox{\ion{Si}{II}} & 7.51$\pm$0.03 & $+$5.35 & $-$2.16 & $+$0.43  \\
\hbox{\ion{K}{I}}   & 5.03$\pm$0.09 & $+$2.98 & $-$2.05 & $+$0.54 \\
\hbox{\ion{Ca}{I}}  & 6.34$\pm$0.04 & $+$4.03 & $-$2.31 & $+$0.27 \\
\hbox{\ion{Ca}{II}} & 6.34$\pm$0.04 & $+$4.43 & $-$1.91 & $+$0.68 \\
\hbox{\ion{Sc}{I}}  & 3.15$\pm$0.04 & $+$0.58 & $-$2.58 & $+$0.01 \\
\hbox{\ion{Sc}{II}} & 3.15$\pm$0.04 & $+$0.75 & $-$2.40 & $+$0.18 \\
\hbox{\ion{Ti}{I}}  & 4.95$\pm$0.05 & $+$2.71 & $-$2.24 & $+$0.34 \\
\hbox{\ion{Ti}{II}} & 4.95$\pm$0.05 & $+$2.69 & $-$2.26 & $+$0.32 \\
\hbox{\ion{V}{I}}   & 3.93$\pm$0.08 & $+$1.44 & $-$2.49 & $+$0.09 \\
\hbox{\ion{V}{II}}  & 3.93$\pm$0.08 & $+$1.70 & $-$2.23 & $+$0.36  \\
\hbox{\ion{Cr}{I}}  & 5.64$\pm$0.04 & $+$2.95 & $-$2.69 & $-$0.11 \\
\hbox{\ion{Cr}{II}} & 5.64$\pm$0.04 & $+$3.32 & $-$2.32 & $+$0.26 \\
\hbox{\ion{Mn}{I}}  & 5.43$\pm$0.04 & $+$2.56 & $-$2.87 & $-$0.29 \\
\hbox{\ion{Co}{I}}  & 4.99$\pm$0.07 & $+$2.69 & $-$2.30 & $+$0.29 \\
\hbox{\ion{Ni}{I}}  & 6.22$\pm$0.04 & $+$3.76 & $-$2.46 & $+$0.12 \\
\hbox{\ion{Ni}{II}} & 6.22$\pm$0.04 & $+$3.88 & $-$2.34 & $+$0.25 \\
\hbox{\ion{Zn}{I}}  & 4.56$\pm$0.05 & $+$2.22 & $-$2.34 & $+$0.25  \\
\hbox{\ion{Sr}{II}} & 2.87$\pm$0.07 & $+$0.10 & $-$2.77 & $-$0.18 \\
\hbox{\ion{Y}{II}}  & 2.21$\pm$0.05 & $-$0.78 & $-$2.99 & $-$0.40 \\
\hbox{\ion{Zr}{II}} & 2.58$\pm$0.04 & $-$0.07 & $-$2.65 & $-$0.07 \\
\hbox{\ion{Ba}{II}} & 2.18$\pm$0.09 & $-$1.22 & $-$3.40 & $-$0.81 \\
\hbox{\ion{La}{II}} & 1.10$\pm$0.04 & $<-$1.85 & $<-$2.95 & $<-$0.36 \\
\hbox{\ion{Ce}{II}} & 1.58$\pm$0.04 & $-$0.83 & $-$2.41 & $+$0.18 \\
\hbox{\ion{Eu}{II}} & 0.52$\pm$0.04 & $-$2.35 & $-$2.87 & $-$0.28 \\
\noalign{\smallskip}
\hline
\end{tabular}
\end{table}

In Table \ref{finalabundNLTE} we present the complete line list analysed in NLTE, 
with the respective individual abundances. Below we also give element-by-element 
information about sources of the NLTE atomic models we used, and we also present 
the graphical results of the NLTE line synthesis. 

\begin{table}
\caption{Isotopic abundance fractions in the solar system from 
Asplund et al. (2009), relevant for HFS computations adopted in 
the present work.}             
\label{fractions}      
\scalefont{1.0}
\centering                          
\begin{tabular}{ccr}        
\hline\hline                 
\noalign{\smallskip}
\hbox{Elements} & \hbox{Isotopes} & \hbox{\%} \\
\noalign{\smallskip}
\hline
\noalign{\smallskip}
\hbox{Sodium}    & \hbox{$^{23}$Na} & 100.000 \\
\hbox{Aluminum}  & \hbox{$^{27}$Al} & 100.000 \\
\hbox{Potassium} & \hbox{$^{39}$K}  & 93.132 \\
\hbox{}          & \hbox{$^{40}$K}  & 0.147 \\
\hbox{}          & \hbox{$^{41}$K}  & 6.721 \\
\hbox{Scandium}  & \hbox{$^{45}$Sc} & 100.000 \\
\hbox{Vanadium}  & \hbox{$^{50}$V}  & 0.250 \\
\hbox{}          & \hbox{$^{51}$V}  & 99.750 \\
\hbox{Manganese} & \hbox{$^{55}$Mn} & 100.000 \\
\hbox{Zinc}      & \hbox{$^{64}$Zn} & 48.630 \\
\hbox{}          & \hbox{$^{66}$Zn} & 27.900 \\
\hbox{}          & \hbox{$^{67}$Zn} & 4.100 \\
\hbox{}          & \hbox{$^{68}$Zn} & 18.750 \\
\hbox{}          & \hbox{$^{70}$Zn} & 0.620 \\
\hbox{Barium}    & \hbox{$^{130}$Ba}& 0.106 \\
\hbox{}          & \hbox{$^{132}$Ba}& 0.101 \\
\hbox{}          & \hbox{$^{134}$Ba}& 2.417 \\
\hbox{}          & \hbox{$^{135}$Ba}& 6.592 \\
\hbox{}          & \hbox{$^{136}$Ba}& 7.854 \\
\hbox{}          & \hbox{$^{137}$Ba}& 11.232 \\
\hbox{}          & \hbox{$^{138}$Ba}& 71.698 \\
\hbox{Lanthanum} & \hbox{$^{138}$La}& 0.091 \\
\hbox{}          & \hbox{$^{139}$La}& 99.909 \\
\hbox{Europium}  & \hbox{$^{151}$Eu}& 47.81 \\
\hbox{}          & \hbox{$^{153}$Eu}& 52.19 \\
\noalign{\smallskip}
\hline
\end{tabular}
\end{table}

\begin{table}
\caption{NLTE abundances derived in HD~140238 using NLTE equivalent 
width fitting (EW) and NLTE line profile fitting (PF).}             
\label{finalabundNLTE}      
\scalefont{1.0}
\centering                          
\begin{tabular}{ccrrr}        
\hline\hline                 
\noalign{\smallskip}
\hbox{Ion} & \hbox{$\lambda$~(\AA)} & \hbox{(X/H)+12} & \hbox{[X/H]} & \hbox{Method} \\
\noalign{\smallskip}
\hline
\noalign{\smallskip}
\hbox{\ion{C}{I}}  & 8335.14 & $+$6.00 & $-$2.43 & PF \\ 
\hbox{\ion{C}{I}}  & 9061.43 & $+$5.97 & $-$2.46 & PF \\
\hbox{\ion{C}{I}}  & 9062.48 & $+$6.03 & $-$2.40 & PF \\
\hbox{\ion{O}{I}}  & 7771.94 & $+$7.04 & $-$1.67 & PF \\
\hbox{\ion{O}{I}}  & 7774.16 & $+$6.98 & $-$1.73 & PF \\
\hbox{\ion{O}{I}}  & 7775.39 & $+$7.00 & $-$1.71 & PF \\
\hbox{\ion{O}{I}}  & 8446.36 & $+$7.04 & $-$1.67 & PF \\
\hbox{\ion{Na}{I}} & 5889.95 & $+$3.37 & $-$2.88 & PF \\
\hbox{\ion{Na}{I}} & 5895.92 & $+$3.37 & $-$2.88 & PF \\
\hbox{\ion{Na}{I}} & 8194.82 & $+$3.37 & $-$2.88 & PF \\
\hbox{\ion{Mg}{I}} & 4167.27 & $+$5.38 & $-$2.16 & PF \\
\hbox{\ion{Mg}{I}} & 4571.10 & $+$5.38 & $-$2.16 & PF \\
\hbox{\ion{Mg}{I}} & 4702.99 & $+$5.38 & $-$2.16 & PF \\
\hbox{\ion{Mg}{I}} & 5172.68 & $+$5.38 & $-$2.16 & PF \\
\hbox{\ion{Mg}{I}} & 5183.60 & $+$5.38 & $-$2.16 & PF \\
\hbox{\ion{Mg}{I}} & 5528.40 & $+$5.38 & $-$2.16 & PF \\
\hbox{\ion{Mg}{I}} & 5711.09 & $+$5.38 & $-$2.16 & PF \\
\hbox{\ion{Mg}{I}} & 8806.76 & $+$5.38 & $-$2.16 & PF \\
\hbox{\ion{Al}{I}} & 3944.01 & $+$3.70 & $-$2.73 & PF \\
\hbox{\ion{Al}{I}} & 3961.52 & $+$3.65 & $-$2.78 & PF \\
\hbox{\ion{K}{I}}  & 7698.96 & $+$2.78 & $-$2.33 & PF \\
\hbox{\ion{Ca}{I}} & 4226.73 & $+$4.04 & $-$2.27 & PF \\
\hbox{\ion{Ca}{I}} & 4283.01 & $+$4.21 & $-$2.10 & EW \\
\hbox{\ion{Ca}{I}} & 4289.37 & $+$4.14 & $-$2.17 & EW \\
\hbox{\ion{Ca}{I}} & 4302.53 & $+$4.15 & $-$2.16 & EW \\
\hbox{\ion{Ca}{I}} & 4318.65 & $+$4.09 & $-$2.22 & EW \\
\hbox{\ion{Ca}{I}} & 4425.44 & $+$4.18 & $-$2.13 & EW \\
\hbox{\ion{Ca}{I}} & 4434.96 & $+$4.07 & $-$2.24 & EW \\
\hbox{\ion{Ca}{I}} & 4435.68 & $+$4.17 & $-$2.14 & EW \\
\hbox{\ion{Ca}{I}} & 4454.78 & $+$4.04 & $-$2.27 & EW \\
\hbox{\ion{Ca}{I}} & 5188.84 & $+$4.20 & $-$2.11 & EW \\
\hbox{\ion{Ca}{I}} & 5261.70 & $+$4.15 & $-$2.16 & EW \\
\hbox{\ion{Ca}{I}} & 5265.56 & $+$4.20 & $-$2.11 & EW \\
\hbox{\ion{Ca}{I}} & 5349.46 & $+$4.22 & $-$2.09 & EW \\
\hbox{\ion{Ca}{I}} & 5512.98 & $+$4.09 & $-$2.22 & EW \\
\hbox{\ion{Ca}{I}} & 5581.96 & $+$4.22 & $-$2.09 & EW \\
\hbox{\ion{Ca}{I}} & 5588.75 & $+$4.11 & $-$2.20 & EW \\
\hbox{\ion{Ca}{I}} & 5590.11 & $+$4.22 & $-$2.09 & EW \\
\hbox{\ion{Ca}{I}} & 5594.46 & $+$4.19 & $-$2.12 & EW \\
\hbox{\ion{Ca}{I}} & 5857.45 & $+$4.17 & $-$2.14 & EW \\
\hbox{\ion{Ca}{I}} & 6102.72 & $+$4.14 & $-$2.17 & EW \\
\hbox{\ion{Ca}{I}} & 6122.22 & $+$4.14 & $-$2.17 & EW \\
\hbox{\ion{Ca}{I}} & 6162.17 & $+$4.05 & $-$2.26 & EW \\
\hbox{\ion{Ca}{I}} & 6169.56 & $+$4.17 & $-$2.14 & EW \\
\hbox{\ion{Ca}{I}} & 6439.07 & $+$4.08 & $-$2.23 & EW \\
\hbox{\ion{Ca}{I}} & 6462.56 & $+$4.06 & $-$2.25 & EW \\
\hbox{\ion{Ca}{I}} & 6471.66 & $+$4.08 & $-$2.23 & EW \\
\hbox{\ion{Ca}{I}} & 6493.78 & $+$4.10 & $-$2.21 & EW \\
\hbox{\ion{Ca}{I}} & 6499.65 & $+$4.23 & $-$2.08 & EW \\
\hbox{\ion{Ca}{I}} & 7148.15 & $+$4.19 & $-$2.12 & EW \\
\hbox{\ion{Ca}{I}} & 7326.14 & $+$4.19 & $-$2.12 & EW \\
\hbox{\ion{Ca}{II}}& 3933.68 & $+$4.15 & $-$2.16 & PF \\
\hbox{\ion{Ca}{II}}& 3968.47 & $+$4.15 & $-$2.16 & PF \\
\hbox{\ion{Ca}{II}}& 8498.02 & $+$4.15 & $-$2.16 & PF \\
\hbox{\ion{Ca}{II}}& 8542.09 & $+$4.15 & $-$2.16 & PF \\
\hbox{\ion{Ca}{II}}& 8662.14 & $+$4.15 & $-$2.16 & PF \\
\hbox{\ion{Sr}{II}}& 4077.72 & $+$0.02 & $-$2.90 & PF \\
\hbox{\ion{Sr}{II}}& 4215.52 & $+$0.07 & $-$2.85 & PF \\
\hbox{\ion{Sr}{II}}&10327.31 & $+$0.00 & $-$2.92 & PF \\
\hbox{\ion{Ba}{II}}& 4554.03 & $-$1.07 & $-$3.24 & PF \\
\hbox{\ion{Ba}{II}}& 6141.70 & $-$1.07 & $-$3.24 & PF \\
\hbox{\ion{Ba}{II}}& 6496.92 & $-$1.01 & $-$3.18 & PF \\
\noalign{\smallskip}
\hline
\end{tabular} 
\end{table}

\subsection{Light elements}

\begin{figure}
\centering
\resizebox{80mm}{!}{\includegraphics[angle=0]{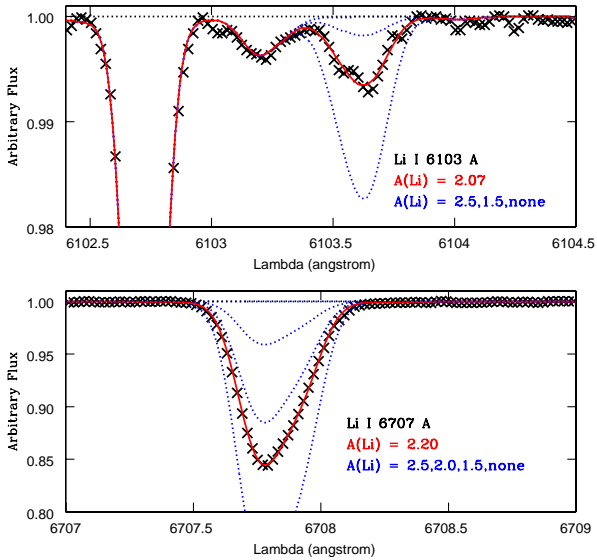}}
\caption{LTE lithium abundance in HD~140283 from the \ion{Li}{I} lines 
located at 6103~{\AA} (upper panel) and 6707~{\AA} (lower panel). 
Observations (crosses) are compared with synthetic spectra computed with 
different abundances (blue dotted lines), as well as with the adopted 
abundances (red solid lines).}
\label{Li_fig}
\end{figure}

$Lithium$. 
We derived the LTE Li abundance A(Li)~$=+2.14$ based on 
the \ion{Li}{I} lines located at 6103~{\AA} and 6707~{\AA} 
(see Fig. \ref{Li_fig}). The wavelength and oscillator strength values 
were adopted from NIST, based on calculations by 
Yan \& Drake (1995). NLTE corrections are given in Asplund et al. (2006) 
for the two Li lines in HD~140283: $+0.09$~dex for 6103~{\AA} and 
$+0.03$ for 6707~{\AA}. The corrected Li abundance 
A(Li)~$=+2.20$ is in excellent agreement with the NLTE 
abundance from Asplund et al. (2006).

$Carbon$. 
In Paper I we analysed the CH A-X electronic transition band (G band) 
in HD~140283, showing a very good agreement between the observed spectrum 
and the computation using the LTE carbon abundance A(C)~$=+6.30$ adopted from 
Honda et al. (2004a). In fact, several CH lines were analysed and all of 
them presented a good fit, leading us to assume that they are properly 
taken into account.

It was also possible to use three \ion{C}{I} lines: 8335.14~{\AA}, 
9061.43~{\AA}, and 9062.48~{\AA}. They are free of telluric and 
other atomic or molecular lines. The lines of \ion{C}{I} in the 
visual part of the spectrum are not detectable. The LTE carbon 
abundance derived from these atomic transitions A(C)~$=+6.44$ is 
slightly higher in comparison with the result from molecular bands. 

The same \ion{C}{I} lines were analysed in NLTE (see upper left panel 
in Fig. \ref{NLTE_O} for the 9062.48~{\AA} line). To perform this work, 
we used the carbon atomic model first proposed by Lyubimkov et al. (2015). 
Our calculations show that NLTE effects are very strong in the analysed lines, 
and act towards a strengthening of their equivalent widths: the ratio between NLTE 
and LTE EWs reaches a factor around 3. This means that simply analysing the program 
carbon lines in LTE approximation results in the derived abundance overestimate 
by about $0.4-0.5$~dex (as it should be for the generally weak lines where equivalent 
width linearly depend upon the number of absorbing atoms). The same conclusion was 
given by Asplund  (2005), who noted that strong NLTE effects in high-excitation neutral 
carbon lines are due to the decrease of source function (compared to the Planck function). 
In particular this is valid for near-infrared \ion{C}{I} lines. Fabbian et al. (2006) 
also predict strong NLTE corrections in the carbon abundances in extremely metal-poor 
stars of  $\sim$$0.4$~dex, in good agreement with the present result.

We find a difference between the C abundance deduced from CH and that 
from \ion{C}{I} lines computed in LTE and NLTE 
of $\Delta($C(CH)~$-$~C(\ion{C}{I})$_{LTE})$~$=-0.14$~dex, and 
$\Delta($C(CH)~$-$~C(\ion{C}{I})$_{NLTE})$~$=+0.30$~dex, respectively. 
The CH lines forming in the upper atmosphere might be affected 
in a 3D modelling calculation (Asplund 2005).

$Nitrogen$. 
The bandhead of CN(0,0) B$_2$$\Sigma$~$-$~X$_2$$\Sigma$ at 3883~{\AA} gives A(N)~$=+6.30$, 
assuming the abundances A(C)~$=+6.30$ and A(O)~$=+7.00$. In Fig. \ref{NO_fig} 
(upper panel) we present the synthetic spectra fitting to the CN bandhead. 
This profile is located in the blue wing of the H8 line of the Balmer series, 
taken into account in the calculations.

$Oxygen$.
The forbidden [\ion{O}{I}]~6300.31~{\AA} line is free from NLTE effects (Kiselman 2001), 
and therefore it is the best oxygen abundance indicator. The line was inspected for telluric 
lines in each of the original exposures. From our series of observations, four of the spectra 
observed in July had to be discarded, given that the telluric lines were masking the oxygen line. 
We then proceeded to co-add the other 19 spectra, and were able to obtain a weak, but measurable line 
(see lower panel in Fig. \ref{NO_fig}). We derive an abundance of A(O)~$=+6.95$, considered in 
this calculation A(C)~$=+6.30$ (C derived from CH lines) and A(N)$=+6.30$.

The triplets at 7771-7775~{\AA} and 8446~{\AA} were also checked to compute the LTE O abundance 
A(O)~$=+7.11$ in HD~140283. Adopting an average of different models presented in Behara et al. (2010) 
for the NLTE corrections to be applied on the triplet 7771-7775~{\AA} lines 
($-0.13$~dex, $-0.10$~dex, and $-0.10$~dex, respectively), we obtain A(O)~$=+6.97$ from this triplet, 
in excellent agreement with the result derived using the forbidden line.

The triplets at 7771-7775~{\AA} (see Fig. \ref{NLTE_O}) and 8446~{\AA} 
were also analysed in NLTE. Our NLTE model of this atom 
was first described in Mishenina et al. (2000), and then updated by 
Korotin et al. (2014). An updated model was applied to study the NLTE 
abundance of this element in cepheids, and its distribution in the 
Galactic disc (see, for instance, Korotin et al. 2014; Martin et al. 2015). 
We obtain a NLTE oxygen abundance A(O)~$=+7.02$ (or [O/Fe]~$=+0.90$), 
in agreement with the derived LTE abundances corrected for NLTE effects.

\subsection{$\alpha$-elements: Mg, Si, S, and Ca}


$Magnesium$. 
After checking the \ion{Mg}{I} lines used by Zhao et al. (1998) in 
the solar spectrum and the bluer lines used by Cayrel et al. (2004) 
for metal-poor stars, we retained 12 lines from an initial list of 21 
profiles to derive the LTE abundance A(\ion{Mg}{I})~$=+5.27$. 
The oscillator strengths were adopted from the critical compilation 
presented in Kelleher \& Podobedova (2008a). The individual abundances 
are consistent among the lines (see upper panels in Fig. \ref{Mg_fig}), 
with the exception of \ion{Mg}{I}~8806.76~{\AA}, which is $0.13$~dex higher 
than the average abundance (see lower panel in Fig. \ref{Mg_fig}).

In addition, it was possible to use two \ion{Mg}{II} lines with the 
spectrum of HD~140283: 4481.13~{\AA} and 4481.15~{\AA}. The derived LTE 
abundance A(\ion{Mg}{II})~$=+5.66$ is higher than the value obtained 
from the neutral species, but both \ion{Mg}{II} lines are blended and 
the result should be used with caution.

Seven magnesium lines were selected for NLTE 
abundance determination. All these lines have very well-defined 
profiles, which provide a robust NLTE abundance of this element (see
Fig. \ref{NLTE_Mg}). To generate \ion{Mg}{I} line profiles, 
we used the NLTE \ion{Mg}{I} atomic model described in detail 
by Mishenina et al. (2004). This model was used in several studies, in 
particular, in the determination of the magnesium abundance in a sample 
of metal-poor stars (Andrievsky et al. 2010).


$Silicon$. 
The main Si abundance indicators are the \ion{Si}{I} lines at 3905.52~{\AA} 
and 4102.94~{\AA}. The line at 3905~{\AA} is blended with CH lines, but these 
molecular lines are weak enough in a rather hot subgiant such that the \ion{Si}{I} line can 
be used. On the other hand, the line at 4102~{\AA} is located in the red wing of the 
H$\delta$ line, and therefore it was necessary to take the hydrogen line 
in the spectrum synthesis into account. Fig. \ref{Si_fig} shows the LTE synthetic spectra 
used for these lines in HD~140283. It was possible to use another three weaker 
\ion{Si}{I} lines located at 5645.61~{\AA}, 5684.48~{\AA}, and 7034.90~{\AA}, with 
individual abundances in good agreement with the previous transitions. 
The final LTE Si abundance obtained was A(\ion{Si}{I})$=+5.30$. 
Because of the good quality of our data, it was possible to derive the LTE silicon abundance 
from the ionized species A(\ion{Si}{II})$=+5.35$, using the \ion{Si}{II}~6371.37~{\AA} line.

$Sulfur$. 
Three \ion{S}{I} lines, which are potentially available for the abundance 
determination in metal-poor stars, are located at 9212.86~{\AA}, 
9228.09~{\AA}, and 9237.54~{\AA}. Unfortunately, the first 
line is severely blended with telluric lines, and the other two lines 
are on a small gap in the spectrum. As a consequence, a reliable sulfur
abundance is not presented here for HD~140283.


$Calcium$. 
To derive the LTE Ca abundance, 36 \ion{Ca}{I} lines were used to obtain 
A(\ion{Ca}{I})$=+4.03$. We adopted the \loggf~values 
from Spite et al. (2012). Because of the high-quality spectrum, even weak and 
blended lines are useful, as shown in Fig. \ref{Ca_fig}. 

We inspected a few \ion{Ca}{II} lines to evaluate the LTE calcium abundance 
from the ionized species. The presence of a strong NLTE effect, however, 
does not permit us to use most of these transitions. 
The LTE abundance A(\ion{Ca}{II})$=+4.43$ was derived based only on the 
\ion{Ca}{II}~8927.36~{\AA} line. We obtained a difference of 
$+0.40$~dex in comparison with the abundance derived from transitions 
of the neutral element. Mashonkina et al. (2007) analysed HD~140283 in 
their study of neutral and singly-ionized calcium in late-type stars, 
and the LTE results show a difference of $+0.30$~dex. Their NLTE 
abundances are [\ion{Ca}{I}/Fe]~$=+0.29\pm0.06$ and [\ion{Ca}{II}/Fe]~$=
+0.24$. Using similar atmospheric parameters for this star, Spite et al. 
(2012) found the NLTE abundances A(\ion{Ca}{I})~$=+4.12\pm0.04$ and 
A(\ion{Ca}{II})~$=+4.08\pm0.05$. Our LTE abundance shows that the 
departure from LTE is not strong for neutral calcium in this star.

The NLTE abundance A(Ca)~$=+4.14$ was determined 
from the average of 35 lines, including: i) the EW analysis of 
30 \ion{Ca}{I} lines with 10~$<$~EW(m{\AA})~$<$~40; 
ii) the profile of five other lines, namely the strongest line of this 
atom at 4226.73~{\AA} (see upper 
panel in Fig. \ref{NLTE_Ca}); the H and K \ion{Ca}{II} lines 
(see lower  panel in Fig. \ref{NLTE_Ca}); and another three 
\ion{Ca}{II} lines located in the red. The NLTE atomic model was 
described in Spite et al. (2012), where it was used for the study of 
halo metal-poor stars.

\subsection{Odd-Z elements: Na, Al, and K}

$Sodium$. 
To derive the LTE sodium abundance we used an initial 
line list based on Baum\"uller et al. (1998) with 
updated oscillator strengths from Kelleher \& Podobedova (2008a). After 
checking 13 \ion{Na}{I} lines, the final result A(\ion{Na}{I})~$=+3.62$ 
is based on five sodium lines. $^{23}$Na is the only stable isotope 
representing the sodium abundance, with nuclear spin I~$=3/2$
\footnote{Adopted from the Particle Data Group (PDG) collaboration: 
http://pdg.lbl.gov/} and therefore exhibiting HFS. The hyperfine 
coupling constants are adopted from Das \& Natarajan (2008) 
and Safronova et al. (1999). When not available, these constants were assumed to be null. 
The HFS for each line were computed by employing a code described and  
made available by McWilliam et al. (2013). Fig. \ref{Na_fig} shows in 
the upper panel the adopted synthetic spectrum for \ion{Na}{I}~5895.92~
{\AA} and \ion{Na}{I}~8194.82~{\AA} lines as example of typical fitting 
procedures.


A NLTE atomic model of this element was presented for the first time in 
Korotin \& Mishenina (1999) and then updated by Dobrovolskas et al. 
(2014). We analysed three \ion{Na}{I} lines: D$_{1}$, D$_{2}$, and the 
line at 8194.82~{\AA}. Synthesized NLTE profiles fitted to the observed 
sodium line profiles are shown in Fig. \ref{NLTE_Na}, and the final 
NLTE abundance A(\ion{Na}{I})~$=+3.37$ was adopted.


$Aluminum$. 
The only stable isotope for aluminum is $^{27}$Al and to derive 
the Al abundance we used the resonance doublet \ion{Al}{I}~3944.01~{\AA} 
and \ion{Al}{I}~3961.52~{\AA}, taking the CH line blending 
with the first line into account. The local continuum around the \ion{Al}{I}~3961.52~{\AA} 
line is defined by the blue wings of the H$\varepsilon$ and H \ion{Ca}{II} lines, 
which were taken into account in the spectrum synthesis. 
$^{27}$Al has nuclear spin I~$=5/2$ and we adopted 
the hyperfine coupling constants from Nakai et al. (2007) and Brown \& Evenson (1999), 
with updated oscillator strengths from Kelleher \& Podobedova (2008b). 
Fig. \ref{Al_K_fig} shows in the upper panel the LTE synthetic 
spectra used for \ion{Al}{I}~3961.52~{\AA} as an example. The adopted LTE 
abundance is A(\ion{Al}{I})$=+2.96$ and must be corrected for NLTE effects.

The NLTE \ion{Al}{I} atomic model presented in Andrievsky et al. (2008) was 
employed and the result of the profile fitting for \ion{Al}{I}~3961.52~{\AA} 
line is shown in Fig. \ref{NLTE_AlK} (upper panel). The final NLTE abundance 
A(\ion{Al}{I})$=+3.68$ is $0.72$~dex higher than the LTE result.

As it was stated above, in some cases the NLTE profile synthesis should 
be combined with the LTE synthesis, which takes the blending lines in 
the vicinity of the studied line into account. For instance, we would get 
the wrong result if we derived the NLTE abundance from UV \ion{Al}{I} lines 
only using the pure MULTI NLTE profiles, since these lines are located in the 
wings of very strong H and K \ion{Ca}{II} lines. 
Therefore it is absolutely necessary to take continuum distortion in their 
vicinity into account. This is made through a combination
of calculations with the codes MULTI NLTE and SYNTHV LTE, which provides 
a correct aluminum abundance.


$Potassium$. 
The final LTE potassium abundance A(\ion{K}{I})~$=+2.98$ was only derived 
from the \ion{K}{I}~7698.96~{\AA} line, as shown in the lower 
panel of Fig. \ref{Al_K_fig}. The other red doublet component at 7665~{\AA} 
is strongly blended with telluric lines and was excluded. For this 
element, we adopted the isotopic abundance fractions in the solar system, 
as described in Asplund et al. (2009) and reproduced in Table 
\ref{fractions}. The HFS was computed taking into account the major 
potassium isotope $^{39}$K, which has nuclear spin I~$=3/2$. We adopted 
hyperfine coupling constants from Belin et al. (1975) and Falke et al. 
(2006), with updated oscillator strength from Sansonetti (2008).

A NLTE \ion{K}{I} atomic model was presented in Andrievsky et al. (2010), 
where it was first employed to derive potassium NLTE abundances in a 
sample of extremely metal-poor halo stars. This model was used to 
synthesize the profile of the \ion{K}{I}~7698.96~{\AA} line in HD 140283
(see lower panel in Fig. \ref{NLTE_AlK}), which gives 
A(\ion{K}{I})~$=+2.78$ as the final abundance.

\subsection{Iron-peak elements: Sc, Ti, V, Cr, Mn, Fe, Co, Ni, and Zn}


$Scandium$. 
We were able to use two \ion{Sc}{I} lines to derive the scandium 
abundance. The lines located at 4020.39~{\AA} and 4023.68~{\AA} present low 
excitation potential and the average abundance A(\ion{Sc}{I})$=+0.58$ should 
suffer from over-ionization via NLTE effects (Zhang et al. 2008). 
$^{45}$Sc is the only stable scandium isotope, with nuclear spin I~$=7/2$,
and to compute the HFS we adopted hyperfine coupling constants from Childs 
(1971), Siefart (1977), and Ba\c{s}ar et al. (2004). On other hand, the 
ionized species presented 19 useful lines, covering a wider range in 
wavelengths and excitation potential values. The average result A(\ion{Sc}{II})
$=+0.75$ is higher in comparison with the abundance from the neutral species 
and it is not affected strongly by NLTE effects, at least at solar metallicity 
(see Asplund et al. 2009). The hyperfine coupling constants were adopted 
from Villemoes et al. (1992), Mansour et al. (1989), and Scott et al. 
(2015). In Fig. \ref{Sc_fig} we show the fitting procedures used for the 
\ion{Sc}{II}~4246.82~{\AA} and \ion{Sc}{II}~5526.79~{\AA} lines as 
examples. 


$Titanium$. 
For titanium, 
we applied equivalent widths to derive the final abundances 
A(\ion{Ti}{I})~$=+2.71$ and A(\ion{Ti}{II})~$=+2.69$ (Sect. \ref{EW}).


$Vanadium$. 
In Paper I we analysed five \ion{V}{I} 
lines and seven \ion{V}{II} lines to derive the vanadium 
abundances A(\ion{V}{I})~$=+1.35\pm0.10$ and A(\ion{V}{II})~$=+1.72\pm0.10$, 
respectively. The average result A(V)~$=+1.56\pm0.11$ is in good 
agreement with A(V)~$=1.55$ from Honda et al. (2004a), derived in their 
analysis from three lines (see Table 3 in Siqueira-Mello et al. 2012 
for details). We adopted seven \ion{V}{I} and eight 
\ion{V}{II} lines, including HFS based on hyperfine coupling 
constants from Unkel et al. (1989), 
Palmeri et al. (1995), Armstrong et al. (2011), Gyzelcimen et al. (2014), 
and Wood et al. (2014), with nuclear spin I~$=7/2$. We only used the major 
V isotope in the computations (see Table \ref{fractions}). 
The oscillator strengths were adopted from Whaling et al. (1985) for \ion{V}
{I} and from Wood et al. (2014) for \ion{V}{II}. We obtained A(\ion{V}{I})~$=
+1.44$ and A(\ion{V}{II})~$=+1.70$, in agreement with the previous results. 
In addition, the final abundances derived from \ion{V}{I} and \ion{V}{II} 
lines are more consistent in the present analysis. 
In Fig. \ref{V_fig} we show the 
\ion{V}{I}~4379.230~{\AA} line (upper panel) and the \ion{V}{II}~4023.378~
{\AA} line (lower panel) as examples. The differences measured between the 
two ionization stages should be explained by strong NLTE effects expected 
for \ion{V}{I} lines.


$Chromium$. 
We derived individual chromium abundances for 28 \ion{Cr}{I} lines,
but we excluded the transitions located at 4756.11~{\AA} and 5237.35~{\AA} 
because of the higher abundances in comparison with results from other lines, 
and we obtained the abundance 
A(\ion{Cr}{I})~$=+2.95$ based on seven \ion{Cr}{I} lines. The fitting 
procedure used for \ion{Cr}{I}~4254.33~{\AA} line is shown in Fig. 
\ref{Cr_Mn_fig}. In addition, it was possible to use seven 
\ion{Cr}{II} lines in HD~140283 to derive the abundance A(\ion{Cr}{II})~$=
+3.32$, higher by 0.37 dex than the result obtained from the neutral species.

$Manganese$. 
The only manganese stable isotope is $^{55}$Mn, 
with nuclear spin I~$=5/2$. 
In addition to the three lines belonging 
to the resonance triplet (\ion{Mn}{I}~4030.75~{\AA}, 4033.06~{\AA}, and 
4034.48~{\AA}), it was also possible to measure 13 subordinate lines.
We took the HFS properly into account based on the hyperfine 
structure line component patterns from Den Hartog et al. (2011), 
which also present the sources for hyperfine coupling constants. 
It is well known that the abundances derived from the triplet
resonance lines are 
systematically lower in comparison with the results from subordinate lines 
in very metal-poor giant stars. Indeed, the result we obtained 
for HD~140283 using the triplet A(\ion{Mn}{I})~$=+2.31$ is lower than 
A(\ion{Mn}{I})~$=+2.56$ derived from the subordinate lines. The resonance 
triplet lines are more susceptible to NLTE effects, 
and for this reason we did not include them in the average
Mn abundance. The line \ion{Mn}{I}~4041.35~{\AA} is 
shown in Fig. \ref{Cr_Mn_fig} (lower panel).

$Iron$.
We used equivalent widths to derive the final 
abundances A(\ion{Fe}{I})~$=+4.91$ and A(\ion{Fe}{II})~$=+4.96$, as described
in Sect. \ref{EW}. 

$Cobalt$.
Individual cobalt abundances were derived from 17 
\ion{Co}{I} lines. The final abundance A(\ion{Co}{I})~$=+2.69$ was 
computed excluding the lines located at 3861.16~{\AA} and 3873.11~{\AA} 
because of the higher differences in comparison with the average abundance. 

$Nickel$. 
For this element 
we derived the abundance A(\ion{Ni}{I})~$=+3.76$ 
based on 43 \ion{Ni}{I} lines. In Fig. \ref{Ni_Zn_fig} (upper panel) we show 
the \ion{Ni}{I}~3858.29~{\AA} line. We were also able to use 
the \ion{Ni}{II}~3769.46~{\AA} line in the present spectrum to derive the 
abundance A(\ion{Ni}{II})~$=+3.88$ in this star. 

$Zinc$. 
The isotopic structure of zinc is complex (see Table \ref{fractions}),  
but HFS is not needed to be accounted for, therefore, we assumed 
Zn as having a unique isotope with wavelengths dominated by the $^{64}$Zn. 
The abundance A(\ion{Zn}{I})~$=+2.21$ was derived from the \ion{Zn}{I}~
4722.15~{\AA} and \ion{Zn}{I}~4810.53~{\AA} (see Fig. \ref{Ni_Zn_fig}, lower 
panel) lines, with oscillator strengths adopted from Roederer \& Lawler (2012).

\subsection{Neutron-capture elements}



\begin{figure}
\centering
\resizebox{80mm}{!}{\includegraphics[angle=0]{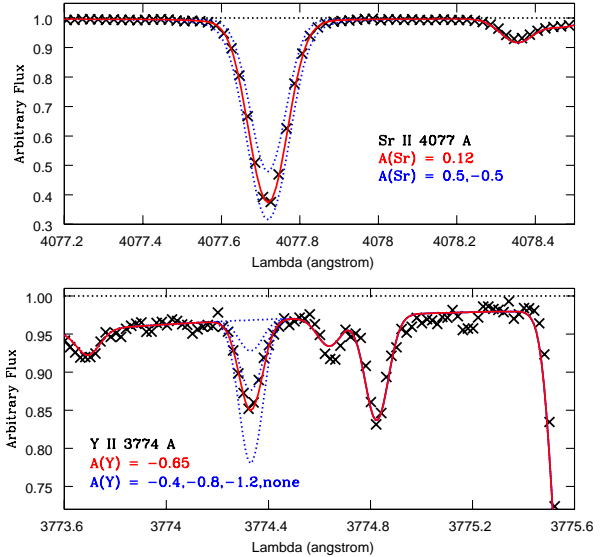}}
\caption{LTE strontium abundance in HD~140283 from \ion{Sr}{II}~4077.72~{\AA} 
line (upper panel) and yttrium abundance from \ion{Y}{II}~3774.33~{\AA} line 
(lower panel). Symbols are the same as in Fig. \ref{Li_fig}.}
\label{Sr_Y_fig}
\end{figure}

$Strontium$. 
In HD~140283 the Sr abundance was derived based on three \ion{Sr}{II} 
lines located at 4077.72~{\AA}, 4215.52~{\AA}, and 10327.31~{\AA}. 
These transitions show HFS effects, but hyperfine coupling constants 
only exist for $^{87}$Sr (Borghs et al. 1983), which accounts for 
less than 7\% of Sr (Asplund et al. 2009). In addition, the atomic lines from 
the even isotopes $^{84}$Sr, $^{86}$Sr, and $^{88}$Sr appear as single lines 
due to the small isotopic splitting for Sr (Hauge 1972). In conclusion, we treated 
each Sr line as a single component, with oscillator strengths adopted from 
Gratton \& Sneden (1994), which gives an average abundance 
of A(\ion{Sr}{II})~$=+0.10$. In the upper panel of Fig. \ref{Sr_Y_fig}, 
we show the fitting procedure used for the \ion{Sr}{II}~4077.72~{\AA} line.

Our NLTE strontium atomic model was described in Andrievsky et al. (2011), 
where it was applied to a sample of metal-poor stars. 
We analysed the same two blue \ion{Sr}{II} lines at 4077.72~{\AA} and 
4215.52~{\AA} (see upper panel in Fig. \ref{NLTE_Sr}), and a third 
line in the near-infrared located at 10327.31~{\AA} (see lower panel in Fig. 
\ref{NLTE_Sr}), with a final NLTE abundance A(\ion{Sr}{II})~$=+0.03$ 
as the adopted result.

In Andrievsky et al. (2011) the NLTE corrections were calculated for the 
\ion{Sr}{II}~4077.72~{\AA}, 4215.52~{\AA}, besides near-infrared lines for different 
temperatures and gravities. Considering Fig. 7 of Andrievsky et al. (2011),
 at \Teff~$=5750$~K and \logg~$=3.7$ the correction 
should be  small and positive. We obtained a NLTE strontium 
abundance that is slightly lower than the LTE abundance. We suggest that
main reasons for this discrepancy can be the result of: a) the LTE results 
from Turbospectrum may use slightly different atomic constants; and b) this 
star is more metal-rich ([Fe/H]~$=-2.59$) than the calculations for [Fe/H]~$=-3.0$ 
given in Andrievsky et al. (2011).

$Yttrium$. 
For yttrium it was possible to check in LTE five \ion{Y}{II} lines, 
using oscillator strengths adopted from Hannaford et al. (1982) and 
Grevesse et al. (2015), with A(\ion{Y}{II})~$=-0.78$ as the final 
abundance. The synthetic profile adopted for \ion{Y}{II}~3774.33~{\AA} 
is shown in Fig. \ref{Sr_Y_fig} (lower panel), which takes the continuum 
affected by the H11 line from the Balmer series into account.

\begin{figure}
\centering
\resizebox{80mm}{!}{\includegraphics[angle=0]{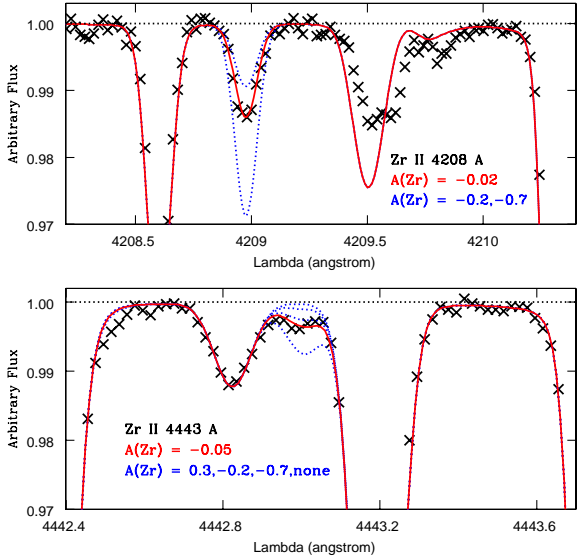}}
\caption{LTE zirconium abundance in HD~140283 from \ion{Zr}{II}~4208.98~{\AA} 
(upper panel) and \ion{Zr}{II}~4443.01~{\AA} (lower panel) lines. 
Symbols are the same as in Fig. \ref{Li_fig}.}
\label{Zr_fig}
\end{figure}

$Zirconium$. 
After inspecting the spectrum, we decided to retain only the three best 
\ion{Zr}{II} lines available, located at 3836.76~{\AA}, 4208.98~{\AA} 
and 4443.01~{\AA}, to derive the LTE zirconium abundance. The \loggf~values 
were adopted from Bi\'emont et al. (1981), with final LTE abundance of 
A(\ion{Zr}{II})~$=-0.07$ (see Fig. \ref{Zr_fig}).

$Barium$. 
The LTE barium abundance was previously analysed in Paper~I, based on 
the \ion{Ba}{II}~4554.03 and \ion{Ba}{II}~4934.08~{\AA} lines, 
with oscillator strengths adopted from Gallagher (1967) and hyperfine structure 
line component patterns from McWilliam (1998). 
We added two other \ion{Ba}{II} lines, located 
at 6141.71~{\AA} and 6496.90~{\AA}, with 
oscillator strengths and HFS from Barbuy et al. (2014). 
With nuclear spin I~$=3/2$, we took the Ba isotopic nuclides into account 
according to Table \ref{fractions}.

\begin{figure}
\centering
\resizebox{80mm}{!}{\includegraphics[angle=0]{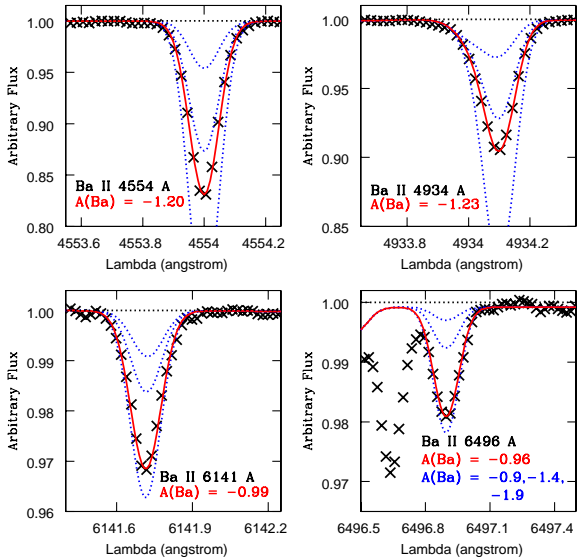}}
\caption{LTE barium abundance in HD~140283 from \ion{Ba}{II}~4554.03~{\AA} 
(upper left panel), \ion{Ba}{II}~4934.08~{\AA} (upper right panel), 
\ion{Ba}{II}~6141.71~{\AA} (lower left panel), and 
\ion{Ba}{II}~6496.90~{\AA} (lower right panel) lines. 
Symbols are the same as in Fig. \ref{Li_fig}.}
\label{Ba_fig}
\end{figure}

Figure \ref{Ba_fig} shows the synthetic profiles computed for the 
\ion{Ba}{II} lines. The blue wing of the \ion{Ba}{II}~4934.08~{\AA} 
line is blended with \ion{Fe}{I}~4934.01~{\AA}, which we took properly 
into account using \loggf~$=-0.59$, adjusted to describe the 
observed spectrum. The individual abundance agrees with the result 
derived from the clear \ion{Ba}{II}~4554.03{\AA} line. To compute the 
profile for \ion{Ba}{II}~6141.71~{\AA}, it is important to include a 
blend with \ion{Fe}{I}~6141.73~{\AA} for which we adopted 
\loggf~$=-1.60$ (Barbuy et al. 2014). For \ion{Ba}{II}~6496.90~
{\AA} there is a telluric line located in the blue wing in the present 
spectrum. The individual abundance agrees with that derived from 
6141.71~{\AA} line, but it is slightly higher in comparison with the 
results from the first two Ba lines. We decided to adopt 
A(\ion{Ba}{II})~$=-1.22$ as the final LTE abundance.

A NLTE atomic model of \ion{Ba}{II} was presented in Andrievsky et al.
(2009). Three \ion{Ba}{II} lines were analysed in HD~140283: 
4554.00~{\AA}, 6141.70~{\AA}, and 6496.92~{\AA} (see Fig. \ref{NLTE_Ba}). 
We applied the odd-to-even isotopic ratio 50:50, which is applicable to old 
metal deficient stars, to synthesize the \ion{Ba}{II}~4554.00~{\AA} line.

\begin{figure}
\centering
\resizebox{80mm}{!}{\includegraphics[angle=0]{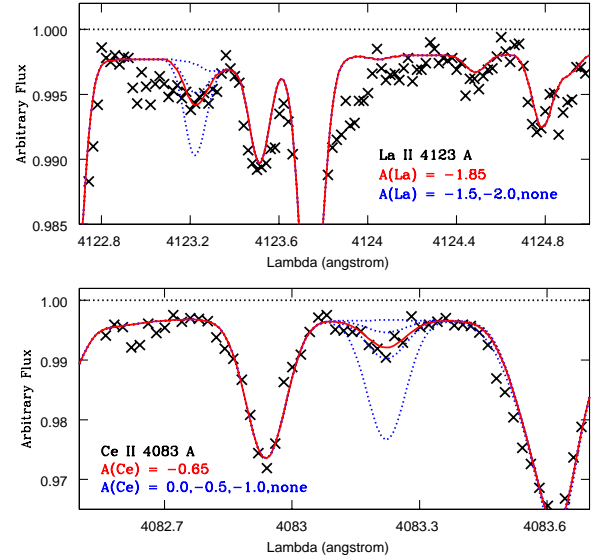}}
\caption{LTE lanthanum abundance in HD~140283 from \ion{La}{II}~4123.22~{\AA} 
line (upper panel) and cerium abundance in from \ion{Ce}{II}~4083.22~{\AA} line 
(lower panel). Symbols are the same as in Fig. \ref{Li_fig}.}
\label{La_Ce_fig}
\end{figure}

$Lanthanum$. 
For lanthanum, available transitions are too weak to enable us to derive 
a robust value for the La abundance, but the upper limit of LTE 
abundance A(\ion{La}{II})~$<-1.85$ was estimated from the 
\ion{La}{II}~4123.22~{\AA} line (see upper panel in Fig. \ref{La_Ce_fig}). 
This profile is located in the red wing of the H$\delta$ line, 
accounted for in the spectrum synthesis. 
We only used the major La isotope, with nuclear spin I~$=7/2$, 
and the hyperfine coupling constants adopted basically from Lawler 
et al. (2001), but also from Furmann et al. (2008) and Honle et al. 
(1982) when not in the basic reference. We also adopted experimental 
oscillator strengths from Lawler et al. (2001).

$Cerium$. 
For cerium, we used the improved laboratory transition probabilities 
presented in Lawler et al. (2009) to derive the LTE abundance 
A(\ion{Ce}{II})~$=-0.83$, based on two \ion{Ce}{II} lines located 
at 4083.22~{\AA} (lower panel in Fig. \ref{La_Ce_fig}) and 4222.60~{\AA}. 
These are well known as good abundance indicators (e.g. Hill et al. 2002). 
The local continuum around \ion{Ce}{II}~4083.22~{\AA} is 
defined by the blue wing of the H$\delta$ line. 
These lines are weak in HD~140283, but still clearly
detectable as a result of the high quality of the spectrum. 
These two lines give [Ce/Fe]~$=+0.36$ and $+0.01$, respectively, 
and a corresponding mean overbundance of cerium [Ce/Fe]~$=+0.18$. 
The overabundance is therefore to be taken with caution.

$Europium$. 
Paper~I was dedicated to derive the LTE europium abundance in HD~140283, 
using the isotopic fractions in the solar material (Table \ref{fractions}) 
with nuclear spin I~$=5/2$. The final abundance A(\ion{Eu}{II})~$=
-2.35\pm0.07$ was obtained from \ion{Eu}{II}~4129.70~{\AA}, which is 
consistent with the upper limit A(\ion{Eu}{II})~$<-2.39$ estimated from 
the \ion{Eu}{II}~4205.05~{\AA} line.

\section{Discussion}

\begin{table}
\caption{Abundances in HD~140283 computed in LTE, NLTE, and 
final abundances adopted. Iron is only computed in LTE.
 }             
\label{adopted_abundances}      
\scalefont{1.0}
\centering                          
\begin{tabular}{c@{}r@{}r@{}r@{}r@{}r}        
\hline\hline                 
\noalign{\smallskip}
\hbox{Element} & \phantom{-}\phantom{-}\hbox{[X/Fe]$_{H04}$} &
\phantom{-}\phantom{-}\hbox{[X/Fe]$_{LTE}$} & 
\phantom{-}\phantom{-}\hbox{[X/Fe]$_{NLTE}$} & 
\phantom{-}\phantom{-}\hbox{[X/Fe]$_{adopted}$}\\
\noalign{\smallskip}
\hline
\noalign{\smallskip}
\hbox{Fe}   & $-$2.53 & $-$2.59 &   ----  & $-$2.59 \\
\hbox{Li}   &    ---- & $+$2.14 & $+$2.20$^\dagger$& $+$2.20  \\
\hbox{C(CH)}& $+$0.47 & $+$0.46 &   ----  & $+$0.46  \\
\hbox{C(CI)}&    ---- & $+$0.60 & $+$0.16 & $+$0.16\\
\hbox{N(CN)}&    ---- & $+$1.06 &  ----   & $+$1.06 \\
\hbox{O}    &    ---- & $+$0.97 & $+$0.90 & $+$0.90\\
\hbox{Na}   &    ---- & $-$0.04 & $-$0.29 & $-$0.29\\
\hbox{Mg}   & $+$0.25 & $+$0.26 & $+$0.43 & $+$0.43\\
\hbox{Al}   & $-$0.94 & $-$0.91 & $-$0.17 & $-$0.17\\
\hbox{Si}   & $+$0.34 & $+$0.38 &    ---- & $+$0.38\\
\hbox{K}    &    ---- & $+$0.54 & $+$0.26 & $+$0.26\\
\hbox{Ca}   & $+$0.33 & $+$0.27 & $+$0.42 & $+$0.42\\
\hbox{Sc}   & $+$0.10 & $+$0.10 &    ---- & $+$0.10\\
\hbox{Ti}   & $+$0.36 & $+$0.33 &    ---- & $+$0.33\\
\hbox{V}    & $+$0.21 & $+$0.22 &    ---- & $+$0.22\\
\hbox{Cr}   & $+$0.30 & $+$0.08 &    ---- & $+$0.08\\
\hbox{Mn}   & $-$0.25 & $-$0.29 &    ---- & $-$0.29\\
\hbox{Co}   & $+$0.25 & $+$0.29 &    ---- & $+$0.29\\
\hbox{Ni}   & $+$0.13 & $+$0.12 &    ---- & $+$0.12\\
\hbox{Zn}   &    ---- & $+$0.25 &    ---- & $+$0.25\\
\hbox{Ge}   &    ---- &    ---- &    ---- &$<-$0.46\\
\hbox{As}   &    ---- &    ---- &    ---- & $+$0.38$^1$\\
\hbox{Se}   &    ---- &    ---- &    ---- & $-$0.15$^1$\\
\hbox{Sr}   & $-$0.31 & $-$0.18 & $-$0.30 & $-$0.30\\
\hbox{Y}    & $-$0.46 & $-$0.40 &    ---- & $-$0.40\\
\hbox{Zr}   & $-$0.14 & $-$0.07 &    ---- & $-$0.07\\
\hbox{Mo}   &    ---- &    ---- &    ---- & $+$0.19$^2$\\
\hbox{Ru}   &    ---- &    ---- &    ---- & $<+$0.99$^2$\\
\hbox{Ba}   & $-$0.96 & $-$0.81 & $-$0.63 & $-$0.63\\
\hbox{La}   &    ---- &$<-$0.36 &   ----  &$<-$0.36 \\
\hbox{Ce}   &    ---- & $+$0.18 &   ----  & $+$0.18\\
\hbox{Eu}   &    ---- & $-$0.28 &   ----  & $-$0.28\\
\hbox{Pt}   &    ---- &   ---- &    ----  & $<+$0.37$^1$\\
\hbox{Pb}   &    ---- &   ---- &    ----  & $<+$1.54$^1$\\
\noalign{\smallskip}
\hline
\end{tabular}
\tablebib{H04: Honda et al. (2004a,b); LTE: this work, in LTE; 
NLTE: this work, in NLTE. $\dagger$: based on the NLTE corrections from Asplund et al. (2006). 
References: 1 Roederer (2012); 2 Peterson (2011).}
\end{table}

\begin{figure}
\centering
\resizebox{80mm}{!}{\includegraphics[angle=0]{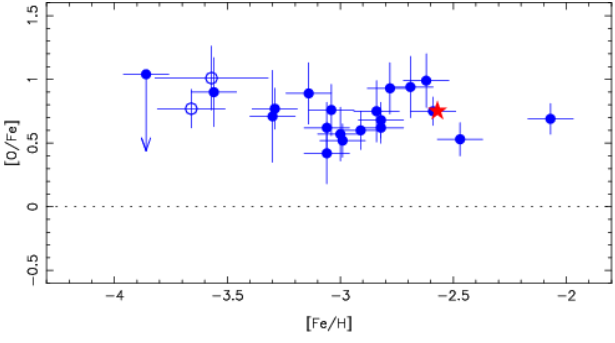}}
\caption{Comparison of the LTE [O/Fe] ratio obtained for the giant stars 
in the Large Programme ``First Stars'' based on the forbidden 
[\ion{O}{I}]~6300.31~{\AA} line (blue filled dots) and in HD~140283 (red star). 
The blue open circles represent the two componnents of the turnoff binary 
CS~22876-032.}
\label{Ocompare_fig}
\end{figure}

\begin{figure}
\centering
\resizebox{80mm}{!}{\includegraphics[angle=0]{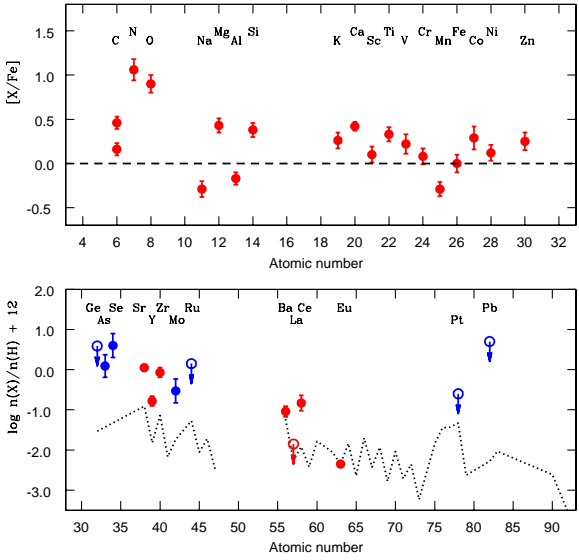}}
\caption{Abundance pattern of HD~140283 based on the present 
results (red circles) and literature (blue circles). 
The dotted line is the abundance pattern of CS~31082-001, 
an r-II EMP star, representing the expected
abundances for an r-rich star, 
rescaled to match the Eu abundance in HD~140283.}
\label{pattern}
\end{figure}

\begin{figure}
\centering
\resizebox{80mm}{!}{\includegraphics[angle=0]{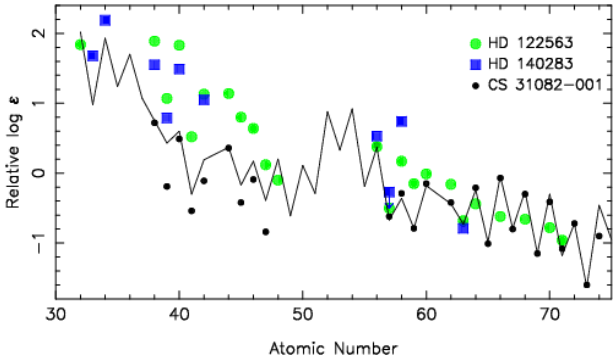}}
\caption{Comparison of the abundance pattern of HD~140283, based on 
the present results and literature (blue squares), with the 
pattern of HD~122563 (green circles) and CS~31082-001 (black dots). 
The solid line is the 
residual solar r-process pattern.}
\label{pattern2}
\end{figure}

In Table \ref{adopted_abundances} we present the abundances 
derived in LTE for 25 elements and an upper limit for lanthanum 
along with abundances derived in NLTE for 9 elements, 
and the final abundances adopted. This table also includes abundances 
from Honda et al. (2004a) and abundances, for As, Se, Pt, Pb, from 
Roederer (2012) and, Mo and Ru, from Peterson (2011). 
All of these are scaled to the metallicity and solar abundances 
adopted in this analysis. 

We adopted as final abundances those analysed with NLTE models 
(Table \ref{finalabundNLTE}), for the nine elements studied 
in NLTE, otherwise we adoted the LTE results. 
The LTE abundances for Sc, V, and Cr are 
the average from neutral and ionized lines. For \ion{Mg}{II}, \ion{Si}{II}, 
and \ion{Ca}{II}, for which individual lines
are reported in Table \ref{linelist},
 the number of useful lines are too small in comparison with 
the neutral species, so we did not include them in the average results.

\subsection{Comparison with the literature}

The forbidden [\ion{O}{I}]~6300.31~{\AA} line is considered 
the best oxygen abundance indicator, given that it is not
affected by NLTE effects. We derive an
abundance of A(O)~$=+6.95$. The triplet lines at 7774-7~{\AA} and 8446~{\AA} 
computed in NLTE give an oxygen abundance A(O)~$=+7.02$, and [O/Fe]~$=+0.90$ 
was derived for HD~140238. Note that Fe was only computed in LTE. 
Mel\'endez et al. (2006) found a difference
of $\Delta$Fe$_{NLTE-LTE}=+0.08$~dex, which would reduce
the oxygen overabundance to [O/Fe]~$=+0.82$. 
This absolute oxygen abundance is in rather good agreement with previous NLTE derivation
by Mel\'endez et al. (2006) of A(O)~$=+6.97$, and [O/Fe]~$=+0.57$, where
the difference in oxygen to iron is due to their
metallicity of [Fe/H]~$=-2.25$ value (in NLTE),
higher by $0.34$~dex than the present value of [Fe/H]~$=-2.59$. 
Values of [Fe/H]~$=-2.3$, [O/H]~$=-1.67$ and [O/Fe]~$=+0.7$ were adopted 
by Bond et al. (2013).

In Fig. \ref{Ocompare_fig} we present a comparison of the 
LTE [O/Fe] abundance ratios obtained for the giant stars 
analysed in the Large Programme ``First Stars'' 
(Cayrel et al. 2004; Spite et al. 2005) based on the 
forbidden [\ion{O}{I}]~6300.31~{\AA} line, 
with the result derived for HD~140283. 
We computed [O/Fe] in HD~140283, adopting the solar oxygen abundance 
A(O)~$=+8.77$ for compatibility with the Large Programme. 
The [O/Fe] ratio obtained in the two components of the well-known 
spectroscopic turnoff binary CS~22876-032 
(Gonz\'alez Hern\'andez et al. 2008) from the OH band, which includes 
corrections for 3D effects, are also presented.

The high nitrogen abundance of [N/Fe]~$=+1.06$, and a consequent 
high [C+N+O/Fe] abundance ratio confirms previous findings by Barbuy (1983). 

The results from Honda et al. (2004a) reported in Table \ref{adopted_abundances} 
were derived from a high-resolution and high-S/N spectrum obtained with the Subaru High 
Dispersion Spectrograph. The Cr abundance is the only result 
significantly different from ours for which the present value is $0.22$~dex lower 
in comparison with their result. In both cases the abundance is the average 
of \ion{Cr}{I} and \ion{Cr}{II} lines. Honda et al. (2004a) adopted a 
\Teff~120~K lower and a gravity \logg~0.2 dex lower with respect to 
our set of parameters, whereas they used the same value for 
microturbulence, which is the most important error source in Cr abundance. However, 
their Cr abundance was derived only based on three lines, in comparison with 
33 lines in the present work, which we consider a more reliable final 
abundance.

Roederer (2012) analysed the zinc abundance in HD 140283 based on 
two \ion{Zn}{I} 
lines also used in the present work, with a spectrum obtained with HARPS/ESO,
 deriving a 
result $0.08$~dex lower in comparison with our abundance. However, the signal-to-noise 
ratio achieved in the present work is much higher than the value from Roederer (2012) 
(250 at 4500~{\AA}), leading us to adopt the present result as the final Zn abundance. 
In addition, Roederer (2012) was able to use the UV \ion{Zn}{II} line located at 
2062.00~{\AA} with a spectrum obtained with STIS/HST to derive A(\ion{Zn}{II})~$=+2.27$, 
or [\ion{Zn}{II}/Fe]~$=+0.30$, which is higher than the result from neutral transitions. 

For strontium, the LTE result of [\ion{Sr}{II}/Fe]~$=-0.17$ derived here agrees with 
the result from Roederer (2012), which used the same two \ion{Sr}{II} lines, 
but with \loggf~$=+0.15$ for \ion{Sr}{II}~4077.7~{\AA}, slightly lower than the adopted value. 
On the other hand, this value is higher in comparison with the LTE results from 
Honda et al. (2004a). Our adopted NLTE analysis gives [\ion{Sr}{II}/Fe]~$=-0.30$. 

The yttrium abundance was derived in Honda et al. (2004a,b) 
based on three \ion{Y}{II} lines,
 located at 3788.70~{\AA}, 3950.35~{\AA}, and 4883.69~{\AA}. 
We added another two \ion{Y}{II} lines and 
the derived abundance [\ion{Y}{II}/Fe]~$=-0.40$ 
is slightly higher than the result from Honda et al. (2004a,b), and it is in 
very good agreement with the abundance derived in Peterson (2011). 

The zirconium abundance we derived is $0.1$~dex higher 
than the result 
from Honda et al. (2004a,b), where two \ion{Zr}{II} lines located 
at 3836.76~{\AA} 
and 4208.98~{\AA} were analysed. We adopted the same values of oscillator 
strengths, and also used an extra \ion{Zr}{II} line located at 4443.01~{\AA}, 
with individual abundances in agreement with the result from the 4208.98~{\AA} 
line. Roederer (2012) also obtained a Zr abundance $0.1$~dex lower than the 
present result, using the \ion{Zr}{II} lines at 
3998.96~{\AA}, 4149.20~{\AA}, and 
4208.98~{\AA}, in common with our analysis. For the latter line, Roederer 
(2012) obtained A(\ion{Zr}{II})~$=-0.09$, only $0.04$~dex lower than the present 
result. In our spectrum the lines 3998.96~{\AA} and 4149.20~{\AA} appear 
severely blended and they were excluded. In addition, Peterson (2011) derived 
an intermediate Zr abundance between the present value and that from Roederer 
(2012), in agreement with our result within error bars.


\subsection{Source of neutron-capture elements}

Truran (1981) suggested an early enrichment of heavy elements uniquely 
through the r-process, arguing that there would be no time for 
the s-process to operate before the formation of the oldest stars. 
In this scenario, even dominantly s-process elements
such as Sr, Y, Zr, and Ba, would have been produced by the r-process in their
relatively reduced amounts. 

Given the recent discussions on the r-process or s-process origin
of the heavy elements in HD~140283, below we try to understand the
results concerning this star. Since HD~140283 is one of the oldest 
stars known so far, formed shortly after the Big Bang, 
it is a natural test star for studies of early heavy element 
formation.

As described in Paper~I, the barium isotopic abundances in HD~140283 are subject 
to intense debate in the literature. In a 1D LTE study of barium isotopes, 
Gallagher et al. (2010, 2012) found  isotope ratios close to the s-process-only 
composition. This is supported by Collet et al. (2009), using 3D hydrodynamical 
models, where a maximum fraction of 15$\pm$34\% contribution of the r-process 
to the isotopic mix in HD~140283 is derived. More recently, Gallagher et al. (2015) 
carried out 3D calculations of the barium isotopic lines in HD~140283, 
and the authors now favour a dominant r-process signature imprinted in the 
barium isotopes in HD~140283. Gallagher et al. (2015) suggest that further work is 
needed to improve the line formation in 3D, and that NLTE has 
to be taken into account.

The [Eu/Ba] ratio is the unambigous indicator as to whether the heavy elements in 
HD~140283 are dominantly r- or s-process. In Paper I we concluded that the 
[Eu/Ba]~$=+0.58\pm0.15$ value indicates that r-process is the dominant nucleosynthesis process. 
We found an LTE Ba abundance A(Ba)~$=-1.22$, in agreement 
within errors with A(Ba)~$=-1.28$ from 1D calculations in LTE derived by 
Gallagher et al. (2015). With the newly derived Ba abundance, the ratio [Eu/Ba]~$=+0.53\pm0.18$ 
confirms the previous conclusion concerning the dominant r-process origin. 

An NLTE Ba abundance A(Ba)~$=-1.05$, or [\ion{Ba}{II}/Fe]~$=-0.63$ is derived here, 
and a lower [Eu/Ba]~$=+0.34$ ratio is obtained in this case. 
If an NLTE Ba abundance is considered, NLTE Eu also has to be considered: 
Mashonkina et al. (2012) presented NLTE abundance corrections for the 
\ion{Eu}{II}~4129~{\AA} line in cool stars, showing that the corrections 
are small (lower than 0.1 dex) and positive for this element and these types of stars, 
which would turn the present [Eu/Ba]$\approx$0.44.
A further ingredient is 3D vs. 1D calculations: Gallagher et al. (2015) 
reported A(Ba)~$=-1.43$ in HD~140283 from LTE 3D calculations, therefore, with 
a $0.15$~dex lower value for the 3D with respect to 1D calculations, 
and in this case LTE [Eu/Ba(3D)]~$=+0.74$ is obtained, or else
adding the 3D effect to [Eu/Ba]$\approx$0.44 above, gives [Eu/Ba]$\approx$0.59.
In Table \ref{eubaratio} we try to summarize all [Eu/Ba] values given above.

According to Simmerer et al. (2004), 
a pure r-process contribution gives [Eu/Ba]$_{r}=+0.698$, whereas 
[Eu/Ba]$_{s}=-1.621$ is the abundance pattern due to the pure s-process. 
Recent models for the solar s-process abundances (Bisterzo et al. 2014) 
predict the same contribution for Ba in comparison with Simmerer et al. (2004), 
and a slightly higher contribution for the Eu abundance: 
2.7\% from Simmerer et al. (2004); 6.0$\pm$0.4\% from Bisterzo et al. (2014). 
In conclusion, the abundance ratios [Eu/Ba] described above do not change 
significantly.

If a  lower [Eu/Ba] value relative to Paper I is confirmed,
it may indicate that the contribution from the 
s-process to the heavy elements is not so small.
 Table \ref{eubaratio} shows that no clear conclusion can be reached,
but that conclusions from Paper I are still favoured.
 This discussion indicates 
that a robust 3D+NLTE synthesis is needed to enable further conclusions.

\begin{table}
\caption{[Eu/Ba] abundance ratios expected from the r- and s-process, 
and derived in HD~140283.}             
\label{eubaratio}      
\scalefont{1.0}
\centering                          
\begin{tabular}{cr}        
\hline\hline                 
\noalign{\smallskip}
\hbox{Source} & \hbox{[Eu/Ba]} \\
\noalign{\smallskip}
\hline
\noalign{\smallskip}
\hbox{$^\dagger$pure r-process} & $+$0.698 \\
\hbox{$^\dagger$pure s-process} & $-$1.621 \\
\hbox{Eu$_{1D+LTE}$~$+$~Ba$_{1D+LTE}$}   & $+$0.53 \\
\hbox{Eu$_{1D+LTE}$~$+$~Ba$_{1D+NLTE}$}  & $+$0.34 \\
\hbox{Eu$_{1D+LTE}$~$+$~$^\triangle$Ba$_{3D+LTE}$}   & $+$0.74 \\
\hbox{$^\diamondsuit$Eu$_{1D+NLTE}$~$+$~Ba$_{1D+NLTE}$} & $+$0.44 \\
\hbox{$^\diamondsuit$Eu$_{1D+NLTE}$~$+$~$^\triangle$Ba$_{3D+NLTE}$} & $+$0.59 \\
\noalign{\smallskip}
\hline
\end{tabular}
\tablebib{$\dagger$: Simmerer et al. (2004); $\triangle$: 3D correction 
from Gallagher et al. (2015); $\diamondsuit$: NLTE correction from 
Mashonkina et al. (2012).}
\end{table}

\subsubsection{Abundance pattern}

Figure \ref{pattern} shows the abundance pattern of HD~140283, based on 
values derived in the present work (red circles), as well as results 
from the literature (blue circles), where upper limits are also indicated. 
The upper panel presents the elements from carbon to zinc, where 
carbon from both CH and \ion{C}{I} lines are indicated. 
The neutron-capture elements are shown in the lower panel. 
As a reference abundance pattern, we used the r-element enhanced EMP (r-II) 
star CS~31082-001 (Barbuy et al. 2011; Siqueira-Mello et al. 2013) 
for comparison (dotted line), rescaled to match the
dominantly r-process europium abundance in HD~140283. 
This figure shows the overabundance of the lighter heavy elements 
(including Sr, Y, and Zr) in comparison with expected values from an r-rich 
star. 

The LTE abundances of Ba and Sr result in [Sr/Ba]~$=+0.63$ for HD~140283. 
If we take the LTE 3D Ba correction from Gallagher et al. (2015) into account, 
this value would be [Sr/Ba]~$=+0.78$, even more overabundant. The newly 
derived NLTE abundances of Ba and Sr give [Sr/Ba]~$=+0.33$, or 
[Sr/Ba]~$=+0.48$ if the 3D Ba correction is applied. 
Overabundances are also obtained for Y and Zr, reaching values of 
[Y/Ba]~$=+0.41$ and [Zr/Ba]~$=+0.74$ in the present work, 
using only the LTE results. With the NLTE Ba abundance these ratios 
decrease to [Y/Ba]~$=+0.23$ and [Zr/Ba]~$=+0.56$ in HD~140283, but the 
3D Ba correction enhances these values to [Y/Ba]~$=+0.38$ and [Zr/Ba]~$=+0.71$. 
In Table \ref{sryzrbaratio} these values are summarized, showing that it is 
important to account for NLTE and 3D effects to evaluate these 
abundance ratios.

\begin{table}
\caption{[Sr, Y, Zr/Ba] abundance ratios derived in HD~140283.}             
\label{sryzrbaratio}      
\scalefont{1.0}
\centering                          
\begin{tabular}{cc}        
\hline\hline                 
\noalign{\smallskip}
\hbox{Source} & \hbox{Abundance} \\
\noalign{\smallskip}
\hline
\noalign{\smallskip}
\hbox{Sr$_{1D+LTE}$~$+$~Ba$_{1D+LTE}$}     & [Sr/Ba]~$=+$0.63 \\
\hbox{Sr$_{1D+NLTE}$~$+$~Ba$_{1D+NLTE}$}   & [Sr/Ba]~$=+$0.33 \\
\hbox{Sr$_{1D+LTE}$~$+$~$^\triangle$Ba$_{3D+LTE}$}     & [Sr/Ba]~$=+$0.78 \\
\hbox{Sr$_{1D+NLTE}$~$+$~$^\triangle$Ba$_{3D+NLTE}$}   & [Sr/Ba]~$=+$0.48 \\
\hbox{Y$_{1D+LTE}$~$+$~Ba$_{1D+LTE}$}      & [Y/Ba]~$=+$0.41 \\
\hbox{Y$_{1D+LTE}$~$+$~Ba$_{1D+NLTE}$}     & [Y/Ba]~$=+$0.23 \\
\hbox{Y$_{1D+LTE}$~$+$~$^\triangle$Ba$_{3D+NLTE}$}     & [Y/Ba]~$=+$0.38 \\
\hbox{Zr$_{1D+LTE}$~$+$~Ba$_{1D+LTE}$}     & [Zr/Ba]~$=+$0.74 \\
\hbox{Zr$_{1D+LTE}$~$+$~Ba$_{1D+NLTE}$}    & [Zr/Ba]~$=+$0.56 \\
\hbox{Zr$_{1D+LTE}$~$+$~$^\triangle$Ba$_{3D+NLTE}$}    & [Zr/Ba]~$=+$0.71 \\
\noalign{\smallskip}
\hline
\end{tabular}
\tablebib{$\triangle$: 3D correction from Gallagher et al. (2015).}
\end{table}

Several authors found high abundance ratios of first peak elements
Sr, Y, Zr, with respect to Ba: [Sr,Y,Zr/Ba]~$>0$, in very metal-poor stars 
(e.g. Honda et al. 2004a,b; Honda et al. 2006; Fran\c{c}ois et al. 2007; 
Cowan et al. 2011). According to Simmerer et al. (2004), in the solar material these 
elements are mainly produced by the s-process, with fractions of 
89\%, 72\%, and 81\%, respectively, but recent results described in 
Bisterzo et al. (2014) are different for some elements (Sr: 68.9$\pm$5.9\%; 
Y: 71.9$\pm$8.0\%; Zr: 66.3$\pm$7.4\%). Therefore, 
while a mechanism responsible for a r-II pattern is claimed to 
explain the 
nucleosynthesis of the heaviest trans-iron elements, an extra 
mechanism (such as truncation or other) should act to produce 
the enhancement of the lightest heavy elements relative to the 
heaviest elements, which must occur very early in the history of the Universe. 

Several models in the literature find that supernova explosion explain the 
overabundances of the first peak elements in metal-poor stars. 
Montes et al. (2007) provided evidence for the existence 
of a light element primary process that contributes to the nucleosynthesis of 
most elements in the Sr to Ag range, producing early high [Sr,Y,Zr/Ba] 
ratios. The astrophysical scenarios in neutrino-driven winds are claimed 
as promising sources of light trans-iron elements (Wanajo 2013, Arcones et al. 
2013). See also the LEPP (e.g. Bisterzo et al. 2014).

Hansen et al. (2014) show that the observed patterns may be obtained 
by combining an r- and an s-pattern, but the 's' has to be (slowly) 
produced in another generation  (AGB?), not very compatible with the great 
age of the star. Yet, a full understanding of core collapse supernova explosion 
using 3D hydrodynamical modelling is needed.


A possibility that HD~140283 formed in a very early dispersed cloud 
could also lead to a previous early chemical enrichment or pollution 
by massive AGB stars, which overproduce the first peak s-process elements 
(Bisterzo et al. 2010). In this scenario, a previous enrichment in Fe seeds is needed, 
and the timescale of the whole process is too long and not 
ideal for such an old star.

In Fig. \ref{pattern2} we show the abundance pattern in HD~140283 
(blue squares) compared with the values obtained in HD~122563 (green circles), 
a metal-poor star well-known for its excesses of light neutron-capture elements. 
The abundances of the heavy elements in HD~122563 have been taken from Honda et al. (2006), 
Cowan et al. (2005), and Roederer et al. (2010). As references, we included 
the abundances of CS~31082-001 (black dots) and the residual solar r-process 
pattern, following the deconvolution by Simmerer et al. (2004). The data were rescaled 
to match the europium abundances. The overabundance 
of first peak elements derived in HD~140283 are only slighly lower than
 the values observed  in HD~122563. 

In conclusion, Figures \ref{pattern} and \ref{pattern2} show overabundances 
of the first-peak heavy elements, and, at a lower level, also of Ba, La and Ce,
 with respect to CS~31082-001 and the residual solar r-process.
It is not clear if the extra mechanism claimed to explain the higher 
abundances of light neutron-capture elements might also produce 
heavier elements, but less efficiently.

The overabundance of cerium [Ce/Fe]~$=+0.18$, a 
dominantly s-process element (83.5$\pm$5.9\% in the solar material, 
from Bisterzo et al. 2014),
was also found in other few stars. 
In Fran\c{c}ois et al. (2007), the star CS~30325-094 also showed a 
high Ce overabundance [Ce/Fe]~$=+0.43$, however, with deficient values 
for other s-elements like Ba ([Ba/Fe]~$=-1.88$) and Sr ([Sr/Fe]~$=-2.24$). 
In Fig. \ref{pattern} and \ref{pattern2} the Ce abundance in 
HD~140283 is clearly higher than the r-process pattern, and the 
same behaviour is observed in HD~122563. Because of the difficulty in explaining 
the abundance pattern of HD~122563 as a combination of 
the r-process and the main s-process, Honda et al. (2006) 
suggested a single process that is responsible for the enhancement of the 
light neutron-capture elements and the production of heavy elements 
in this star. A truncation process in the initial supernova 
(Aoki et al. 2013) could be a possible solution. A new model of hypernova 
shows that the explosion correctly produces the abundances of the elements 
observed in HD~122563, therefore, explaining the so-called weak r-process 
(Fujibayashi et al. 2015) as well as the similar pattern derived in HD~140283.

\section{Conclusions}

We used a high-S/N and high-resolution spectrum of HD~140283, obtained
with a seven hour exposure with the ESPaDOnS spectrograph at the CFHT telescope,
to provide a line list for metal-poor subgiant stars of which HD~140283
is a template. We carried out a detailed derivation of abundances, using
both LTE and NLTE calculations, based on as many as possible reliable lines 
available.

In Paper I we concluded that the derived europium abundance was indicative
of an r-process origin for europium. The present LTE [Eu/Ba]~$=+0.53\pm0.18$ confirms 
that conclusion. Combining the newly derived NLTE Ba abundance with NLTE 
corrections for Eu and 3D corrections for Ba from recent literature, the 
abundance ratio [Eu/Ba]~$=+0.59\pm0.18$ also indicates a small contribution (if any) 
from the main s-process to the neutron-capture elements in HD~140283.

An extra mechanism is claimed to explain the overabundances 
of lighter heavy elements, in addition to an r-II abundance pattern 
responsible for the heavier elements, and possible astrophysical scenarios are 
discussed. The high Ba, La, and Ce abundances derived 
in HD 140283 are similar to those in HD 122563,
 and these two stars may be  excellent examples of abundances dominated 
by the weak r-process.




\begin{acknowledgements}
Based on observations obtained with Brazilian time, provided by
a contract of the Laborat\'orio Nacional de Astrof\'{\i}sica (LNA/MCTI)
and the Canada-France-Hawaii Telescope (CFHT), which is operated by the 
National Research Council of Canada, the Institut National des Sciences 
de l'Univers of the Centre National de la Recherche Scientifique of France,  
and the University of Hawaii.
CS and BB acknowledge grants from CAPES, CNPq, and FAPESP. 
SMA is thankful to FAPESP for financial support and IAG for 
hospitality during his visit to Universidade de S\~ao Paulo. 
MS and FS acknowledge the support of CNRS (PNCG and PNPS). 
SAK acknowledges the SCOPES grant No. IZ73Z0-152485 for financial support. 
We thank the referee for his/her useful comments. 
This work has made use of the VALD database, operated at 
Uppsala University, the Institute of Astronomy RAS in Moscow, 
and the University of Vienna.
\end{acknowledgements}


\begin{appendix} 
\section{Line lists}

\scalefont{1.0}
\begin{longtable}{cccrrcccc}
\caption{\label{EW_measurements} 
Fe and Ti lines adopted: 
species, wavelength ({\rm \AA}), excitation potential (eV),
oscillator strengths from VALD and NIST, measured equivalent widths 
EWs (m{\rm \AA}), errors on the EWs, resulting abundances and associated errors. }\\
\hbox{Species} &  \hbox{$\lambda$({\rm\AA})} & \hbox{$\chi$$_{ex}$(eV)} 
& \hbox{log $gf_{VALD}$} & \hbox{log $gf_{NIST}$} & \hbox{EW(m\AA)} & \hbox{$\sigma$EW(m\AA)} 
& \hbox{A(X)} & \hbox{$\sigma$A(X)} \\
\hline
\endfirsthead
\caption{continued.}\\
\hline\hline
\hbox{Species} &  \hbox{$\lambda$({\rm\AA})} & \hbox{$\chi$$_{ex}$(eV)} 
& \hbox{log $gf_{VALD}$} & \hbox{log $gf_{NIST}$} & \hbox{EW(m\AA)} & \hbox{$\sigma$EW(m\AA)} 
& \hbox{A(X)} & \hbox{$\sigma$A(X)} \\
\hline
\endhead
\hline
\endfoot
\hbox{\ion{Ti}{I}}  & 3989.757 & 0.021 & $-$0.198 & $-$0.126 &  21.37 & 0.73 & 2.73 & 0.02 \\
\hbox{\ion{Ti}{I}}  & 3998.637 & 0.048 & $-$0.056 &    0.016 &  24.37 & 0.40 & 2.69 & 0.01 \\
\hbox{\ion{Ti}{I}}  & 4533.240 & 0.848 &    0.476 &    0.476 &  16.57 & 0.39 & 2.69 & 0.01 \\
\hbox{\ion{Ti}{I}}  & 4534.776 & 0.836 &    0.280 &    0.280 &  11.62 & 0.15 & 2.69 & 0.01 \\
\hbox{\ion{Ti}{I}}  & 4656.468 & 0.000 & $-$1.345 & $-$1.283 &   2.41 & 0.10 & 2.74 & 0.02 \\
\hbox{\ion{Ti}{I}}  & 4681.908 & 0.048 & $-$1.071 & $-$1.009 &   3.66 & 0.15 & 2.70 & 0.02 \\
\hbox{\ion{Ti}{I}}  & 4981.731 & 0.848 &    0.504 &    0.504 &  18.61 & 0.25 & 2.70 & 0.01 \\
\hbox{\ion{Ti}{I}}  & 4991.067 & 0.836 &    0.380 &    0.380 &  15.63 & 0.23 & 2.71 & 0.01 \\
\end{longtable}

\newpage

\scalefont{1.0}
\begin{longtable}{cccrrrr}
\caption{\label{linelist} Line list used in the present analysis:
species, wavelength ({\rm \AA}), oscillator strengths, 
excitation potential (eV), and resulting abundances line by line.}\\
\hline\hline
\hbox{Species} &  \hbox{$\lambda$({\rm\AA})} & \hbox{$\chi$$_{ex}$(eV)}&
\hbox{log $gf$} & \hbox{A(X)} & \hbox{[X/H]} & \hbox{[X/Fe]} \\
\hline
\endfirsthead
\caption{continued.}\\
\hline\hline
\hbox{Species} &  \hbox{$\lambda$({\rm\AA})} & \hbox{$\chi$$_{ex}$(eV)}&
\hbox{log $gf$}  & \hbox{A(X)} & \hbox{[X/H]} & \hbox{[X/Fe]} \\
\hline
\endhead
\hline
\endfoot
\hbox{\ion{Li}{I}}  & 6103.538 & 1.848 &    0.101 & $+$2.07 & $+$1.02 & $+$3.61 \\
\hbox{\ion{Li}{I}}  & 6707.761 & 0.000 & $-$0.009 & $+$2.20 & $+$1.15 & $+$3.74 \\
\hbox{\ion{C}{I}}   & 8335.148 & 7.685 & $-$0.420 & $+$6.47 & $-$1.96 & $+$0.63 \\
\hbox{\ion{C}{I}}   & 9061.433 & 7.483 & $-$0.346 & $+$6.42 & $-$2.01 & $+$0.58 \\
\hbox{\ion{C}{I}}   & 9062.487 & 7.480 & $-$0.456 & $+$6.44 & $-$1.99 & $+$0.60 \\
\hbox{[\ion{O}{I}]} & 6300.311 & 0.000 & $-$9.716 & $+$6.95 & $-$1.74 & $+$0.85 \\
\hbox{\ion{O}{I}}   & 7771.941 & 9.146 &    0.369 & $+$7.12 & $-$1.57 & $+$1.02 \\
\hbox{\ion{O}{I}}   & 7774.161 & 9.146 &    0.223 & $+$7.06 & $-$1.63 & $+$0.96 \\
\end{longtable}

\scalefont{1.0}
\begin{longtable}{cccrrccrrc}
\caption{\label{HFSconstants} Hyperfine coupling constants for the lines 
retained in this analysis.}\\
\hline\hline
\hbox{} &  \hbox{} & \multicolumn{4}{|c|}{Lower level} & \multicolumn{4}{|c|}{Upper level} \\
\hbox{Iso.} &  \hbox{Lines ({\rm\AA})} & \hbox{J} & \hbox{A (MHz)} & \hbox{B (MHz)} 
& \hbox{ref.} & \hbox{J} & \hbox{A (MHz)} & \hbox{B (MHz)} & \hbox{ref.} \\
\hline
\endfirsthead
\caption{continued.}\\
\hline\hline
\hbox{} &  \hbox{} & \multicolumn{4}{|c|}{Lower level} & \multicolumn{4}{|c|}{Upper level} \\
\hbox{Iso.} &  \hbox{Lines ({\rm\AA})} & \hbox{J} & \hbox{A (MHz)} & \hbox{B (MHz)} 
& \hbox{ref.} & \hbox{J} & \hbox{A (MHz)} & \hbox{B (MHz)} & \hbox{ref.} \\
\hline
\endhead
\hline
\endfoot
\hbox{$^{23}$\ion{Na}{I}}  & 4982.813 & $3/2$ & 18.53     & 2.721     & 1 & $5/2$ &          &       &  \\
\hbox{$^{23}$\ion{Na}{I}}  & 5682.633 & $3/2$ & 18.53     & 2.721     & 1 & $3/2$ &          &       &  \\
\hbox{$^{23}$\ion{Na}{I}}  & 5889.950 & $1/2$ & 888.3     &           & 2 & $3/2$ & 18.53    & 2.721 & 1 \\
\hbox{$^{23}$\ion{Na}{I}}  & 5895.924 & $1/2$ & 888.3     &           & 2 & $1/2$ & 94.3485  &       & 1 \\
\hbox{$^{23}$\ion{Na}{I}}  & 8194.824 & $3/2$ & 18.53     & 2.721     & 1 & $5/2$ &          &       &  \\
\hbox{$^{27}$\ion{Al}{I}}  & 3944.006 & $1/2$ & 502.0346  &           & 3 & $1/2$ & 431.84   &       & 4 \\
\hbox{$^{27}$\ion{Al}{I}}  & 3961.520 & $3/2$ & 94.27723  & 18.91898  & 3 & $1/2$ & 431.84   &       & 4 \\
\hbox{$^{39}$\ion{K}{I}}   & 7698.965 & $1/2$ & 230.85986 &           & 5 & $1/2$ & 27.78    &       & 6 \\
\end{longtable}
\tablebib{1: Das \& Natarajan (2008); 2: Safronova et al. (1999); 3: Brown \& Evenson (1999); 
4: Nakai et al. (2007); 5: Belin et al. (1975); 6: Falke et al. (2006); 7: Childs (1971); 
8: Ba\c{s}ar et al. (2004); 9: Villemoes et al. (1992); 10: Mansour et al. (1989); 
11: Scott et al. (2015); 12: Unkel et al. (1989); 13: Gyzelcimen et al. (2014); 
14: Palmeri et al. (1995); 15: Wood et al. (2014); 16: Armstrong et al. (2011); 
17: Lawler et al. (2001).}

\end{appendix}

\begin{appendix} 
\section{Figures of spectral synthesis}

\begin{figure}
\centering
\resizebox{80mm}{!}{\includegraphics[angle=0]{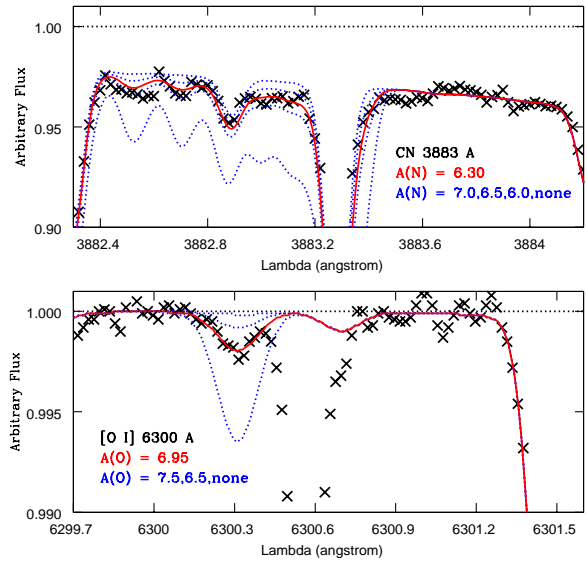}}
\caption{LTE nitrogen abundance in HD~140283 from CN(0,0) 
bandhead at 3883~{\AA} (upper panel) and 
LTE oxygen abundance from [\ion{O}{I}]~6300.31~{\AA} line 
(lower panel). Symbols are the same as in Fig. \ref{Li_fig}.}
\label{NO_fig}
\end{figure}

\begin{figure}
\centering
\resizebox{80mm}{!}{\includegraphics[angle=0]{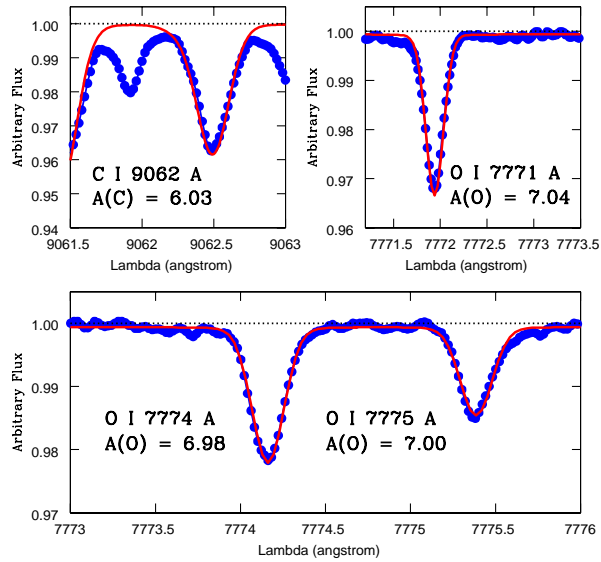}}
\caption{NLTE carbon abundance in HD~140283 from 
\ion{C}{I}~9062.48~{\AA} line (upper left panel) 
and NLTE oxygen abundance from the red \ion{O}{I} triplet: 
7771.94~{\AA} (upper right panel), 7774.16~{\AA}, and 7775.39~{\AA} 
(lower panel) lines. Observations 
(blue circles) are compared with synthetic 
spectra computed with the adopted abundances (red solid lines).}
\label{NLTE_O}
\end{figure}

\begin{figure}
\centering
\resizebox{80mm}{!}{\includegraphics[angle=0]{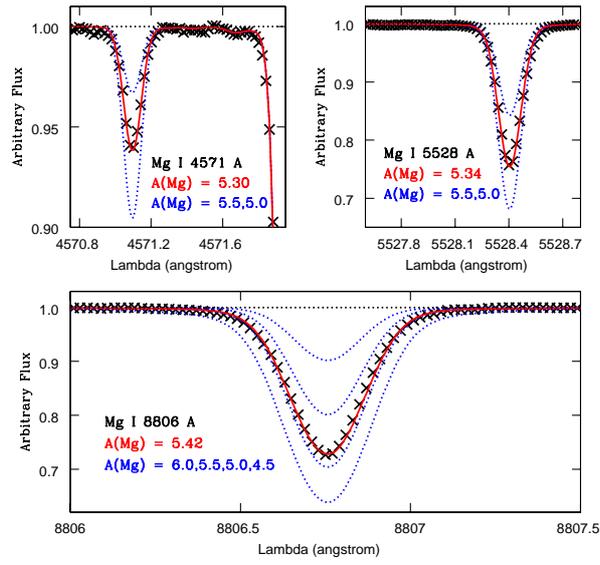}}
\caption{LTE magnesium abundance in HD~140283 from \ion{Mg}{I}~4571.10~{\AA} 
(upper left panel), \ion{Mg}{I}~5528.40~{\AA} (upper right panel), 
and \ion{Mg}{I}~8806.76~{\AA} (lower panel) lines. 
Symbols are the same as in Fig. \ref{Li_fig}.}
\label{Mg_fig}
\end{figure}

\begin{figure}
\centering
\resizebox{80mm}{!}{\includegraphics[angle=0]{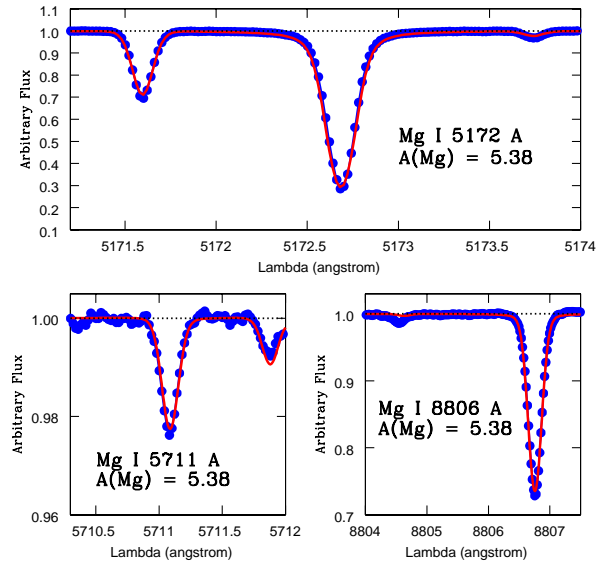}}
\caption{NLTE magnesium abundance in HD~140283 from \ion{Mg}{I}~5172.68~{\AA} 
(upper panel), \ion{Mg}{I}~5711.09~{\AA} (lower left panel), and 
\ion{Mg}{I}~8806.76~{\AA} (lower right panel) lines. Symbols are the same 
as in Fig. \ref{NLTE_O}.}
\label{NLTE_Mg}
\end{figure}

\begin{figure}
\centering
\resizebox{80mm}{!}{\includegraphics[angle=0]{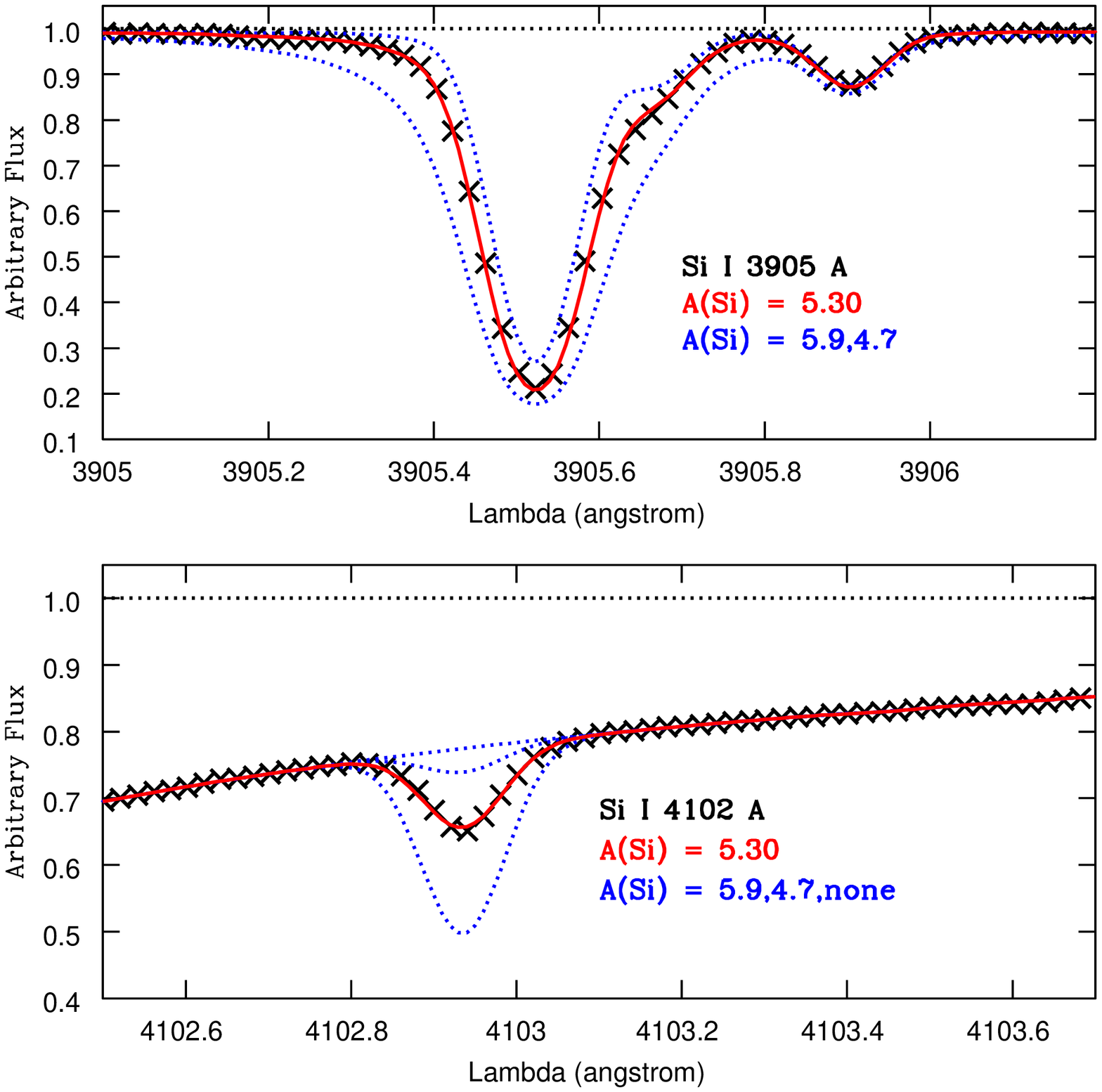}}
\caption{LTE silicon abundance in HD~140283 from \ion{Si}{I}~3905.52~{\AA} 
(upper panel) and \ion{Si}{I}~4102.936~{\AA} (lower panel) lines. 
Symbols are the same as in Fig. \ref{Li_fig}.}
\label{Si_fig}
\end{figure}

\begin{figure}
\centering
\resizebox{80mm}{!}{\includegraphics[angle=0]{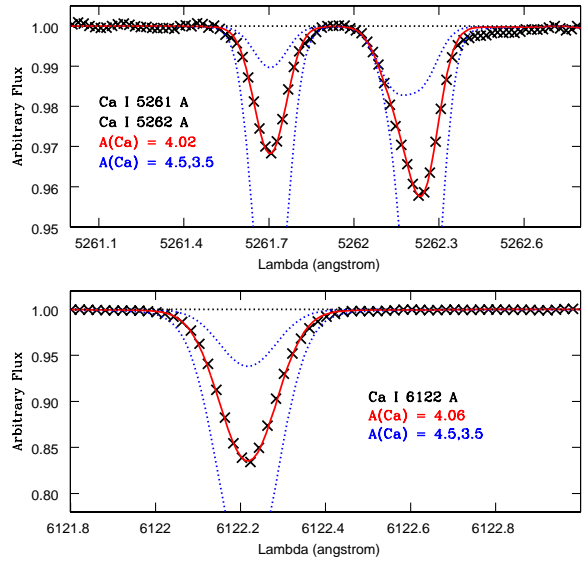}}
\caption{LTE calcium abundance in HD~140283 from \ion{Ca}{I}~5261.70~{\AA}, 
\ion{Ca}{I}~5262.24~{\AA} (upper panel), and \ion{Ca}{I}~6122.23~{\AA} (lower panel) lines. 
Symbols are the same as in Fig. \ref{Li_fig}.}
\label{Ca_fig}
\end{figure}

\begin{figure}
\centering
\resizebox{80mm}{!}{\includegraphics[angle=0]{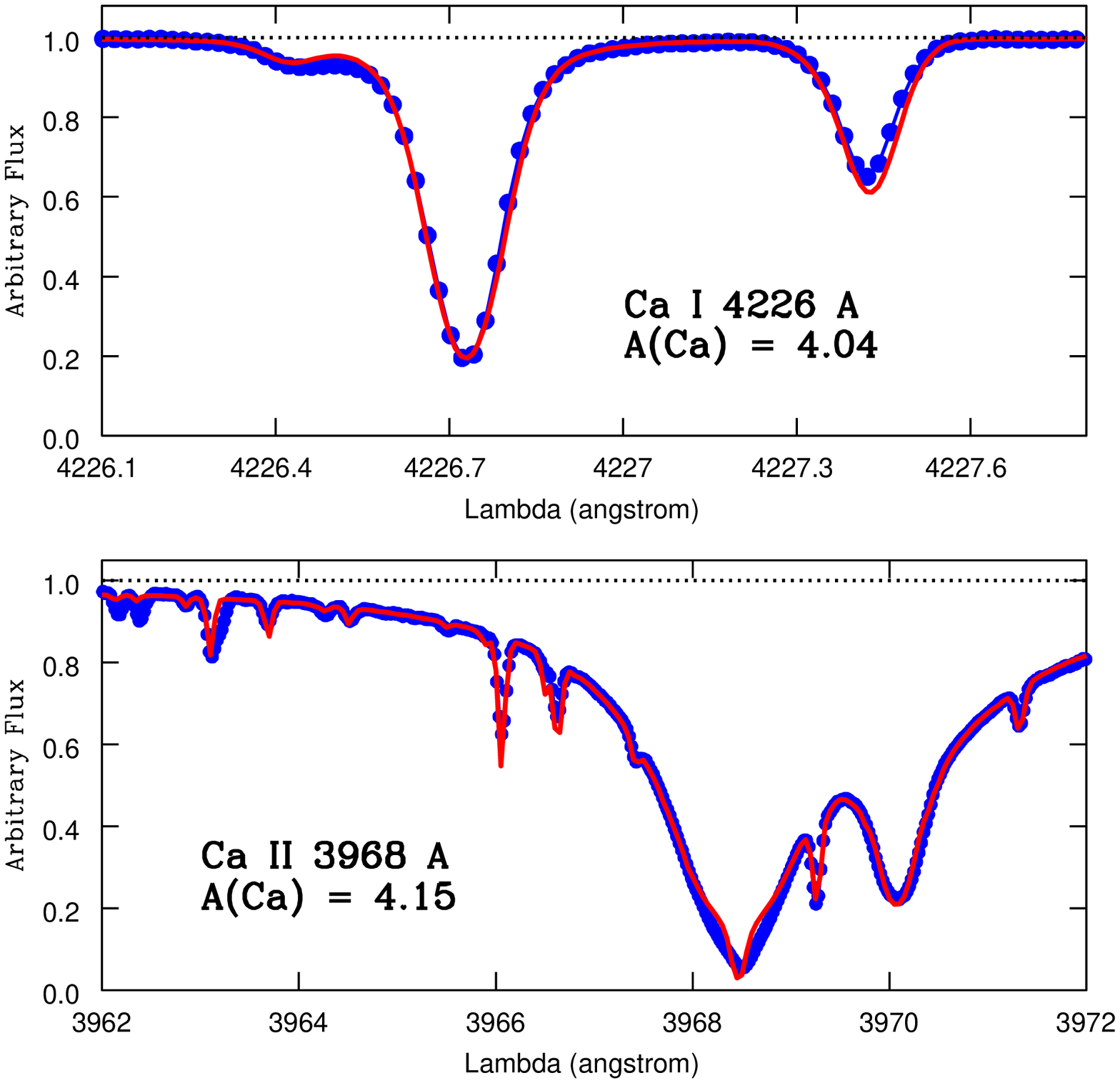}}
\caption{NLTE calcium abundance in HD~140283 from \ion{Ca}{I}~4226.73~{\AA} 
(upper panel) and \ion{Ca}{II}~3968.47~{\AA} (lower panel) lines. 
Symbols are the same as in Fig. \ref{NLTE_O}.}
\label{NLTE_Ca}
\end{figure}

\begin{figure}
\centering
\resizebox{80mm}{!}{\includegraphics[angle=0]{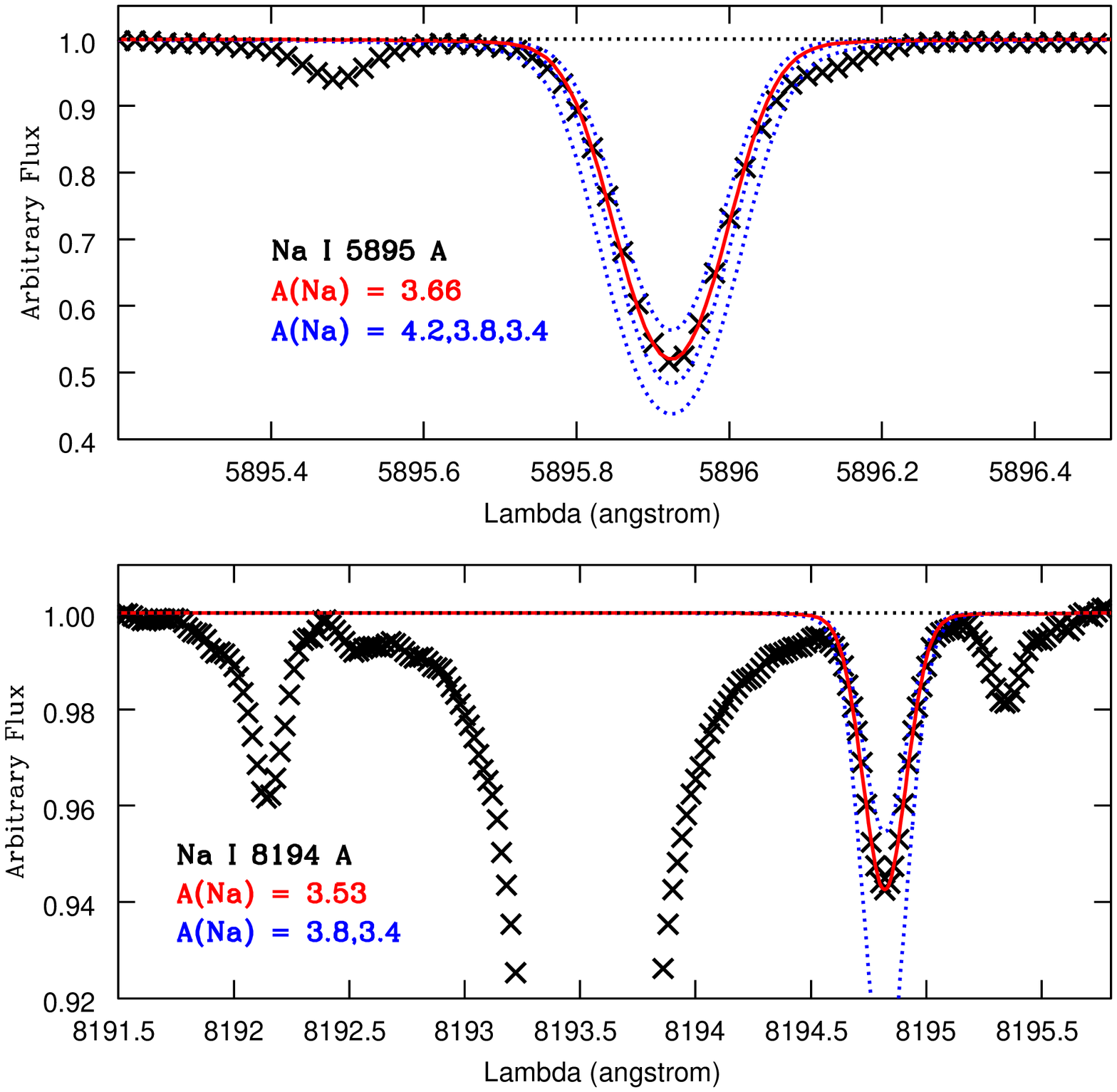}}
\caption{LTE sodium abundance in HD~140283 from \ion{Na}{I}~5895.92~{\AA} 
(upper panel) and \ion{Na}{I}~8194.82~{\AA} (lower panel) lines. 
Symbols are the same as in Fig. \ref{Li_fig}.}
\label{Na_fig}
\end{figure}

\begin{figure}
\centering
\resizebox{80mm}{!}{\includegraphics[angle=0]{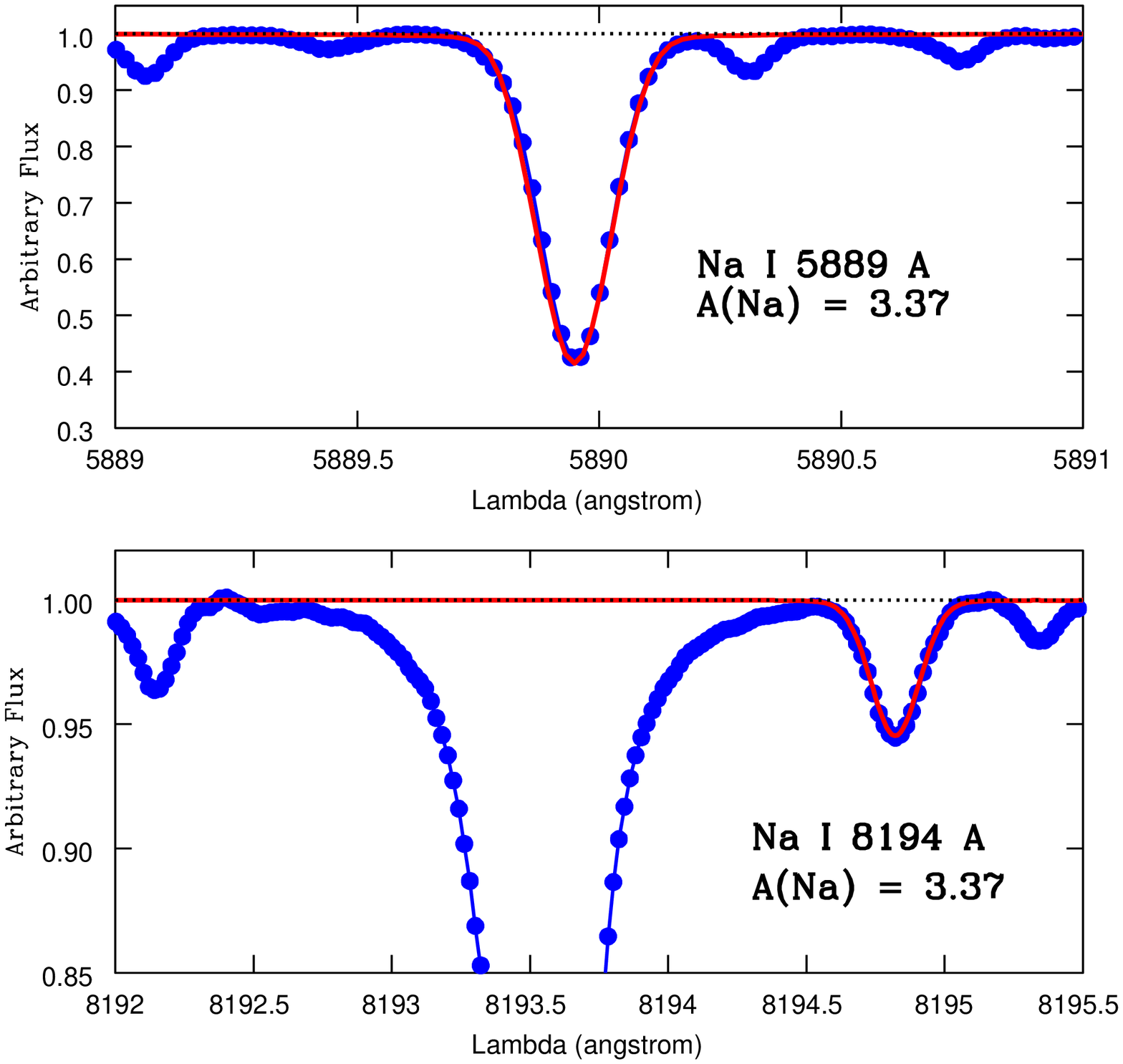}}
\caption{NLTE sodium abundance in HD~140283 from \ion{Na}{I}~5889.95~
{\AA} (D$_{2}$) (upper panel) and \ion{Na}{I}~8194.82~{\AA} (lower panel) 
lines. Symbols are the same as in Fig. \ref{NLTE_O}.}
\label{NLTE_Na}
\end{figure}

\begin{figure}
\centering
\resizebox{80mm}{!}{\includegraphics[angle=0]{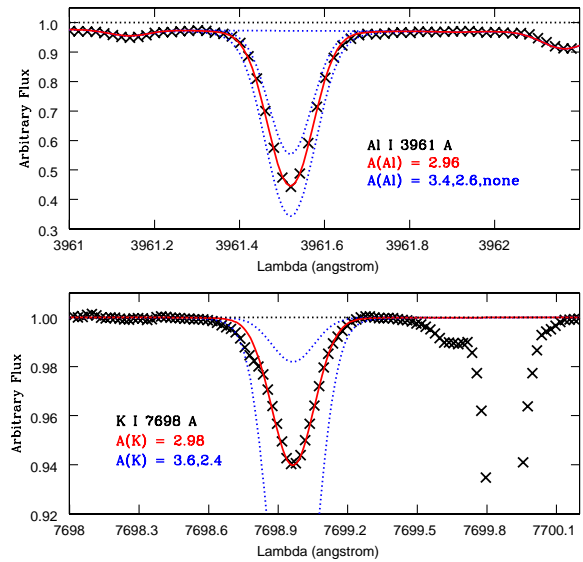}}
\caption{LTE aluminum abundance in HD~140283 from \ion{Al}{I}~3961.52~
{\AA} line (upper panel) and potassium abundance from \ion{K}{I}~7698.96~
{\AA} line (lower panel). Symbols are the same as in Fig. \ref{Li_fig}.}
\label{Al_K_fig}
\end{figure}

\begin{figure}
\centering
\resizebox{80mm}{!}{\includegraphics[angle=0]{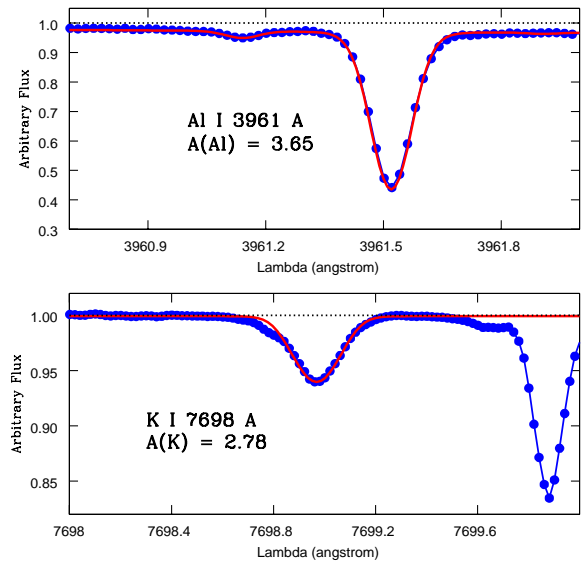}}
\caption{NLTE aluminum abundance in HD~140283 from \ion{Al}{I}~3961.52~{\AA} 
(upper panel) and NLTE potassium abundance from the \ion{K}{I}~7698.96~{\AA} 
line (lower panel) lines. 
Symbols are the same as in Fig. \ref{NLTE_O}.}
\label{NLTE_AlK}
\end{figure}

\begin{figure}
\centering
\resizebox{80mm}{!}{\includegraphics[angle=0]{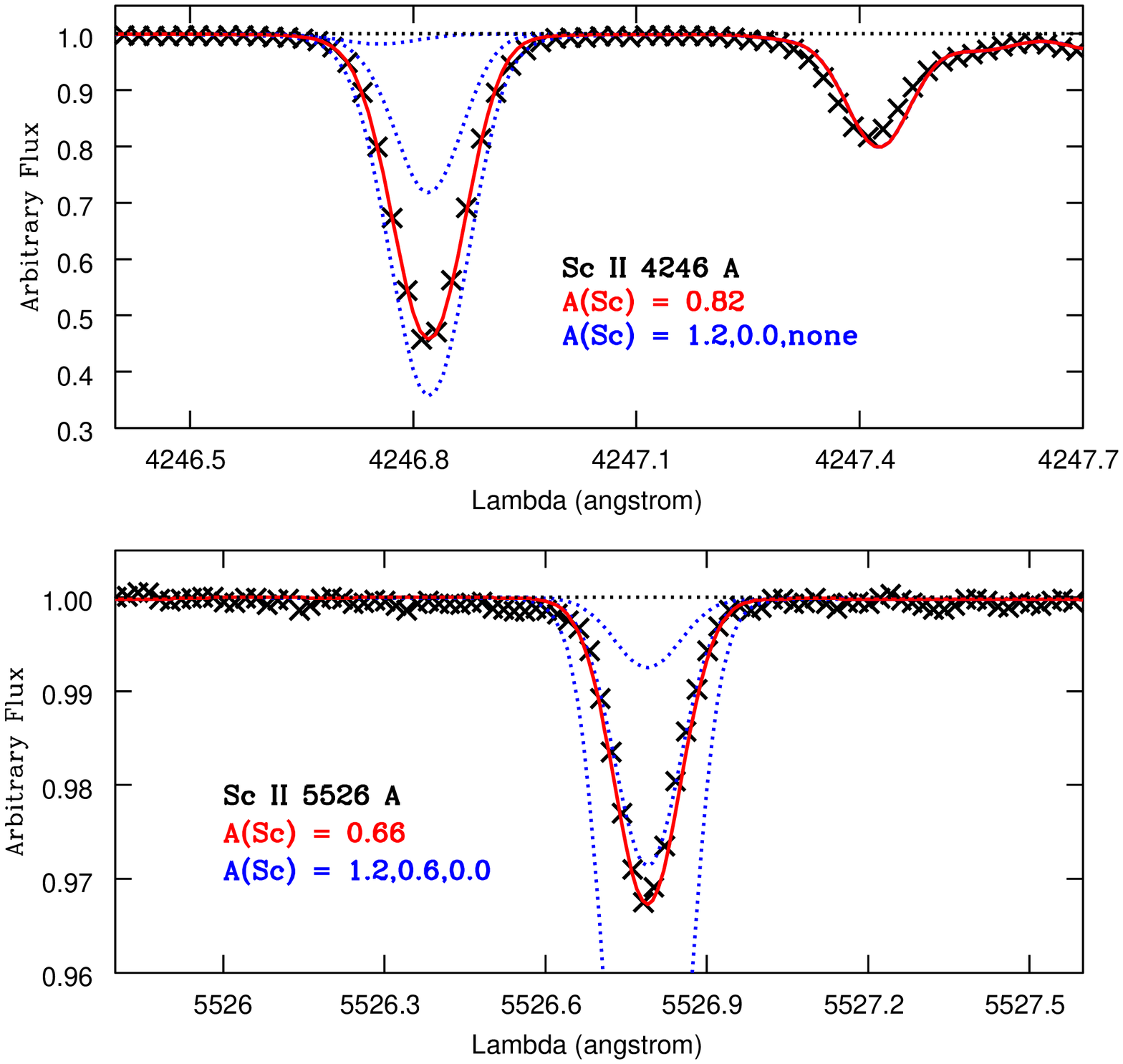}}
\caption{LTE scandium abundances in HD~140283 from \ion{Sc}{II}~4246.82~
{\AA} (upper panel) and \ion{Sc}{II}~5526.79~{\AA} (lower panel) lines. 
Symbols are the same as in Fig. \ref{Li_fig}.}
\label{Sc_fig}
\end{figure}

\begin{figure}
\centering
\resizebox{80mm}{!}{\includegraphics[angle=0]{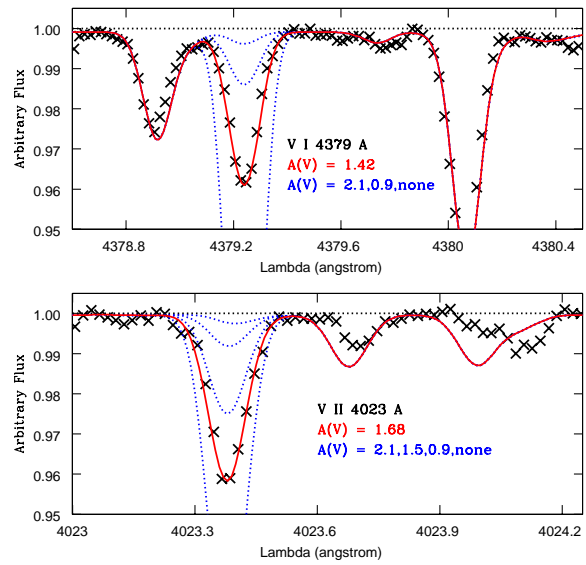}}
\caption{LTE vanadium abundances in HD~140283 from \ion{V}{I}~4379~{\AA} 
(upper panel) and \ion{V}{II}~4023~{\AA} (lower panel) lines. 
Symbols are the same as in Fig. \ref{Li_fig}.}
\label{V_fig}
\end{figure}

\begin{figure}
\centering
\resizebox{80mm}{!}{\includegraphics[angle=0]{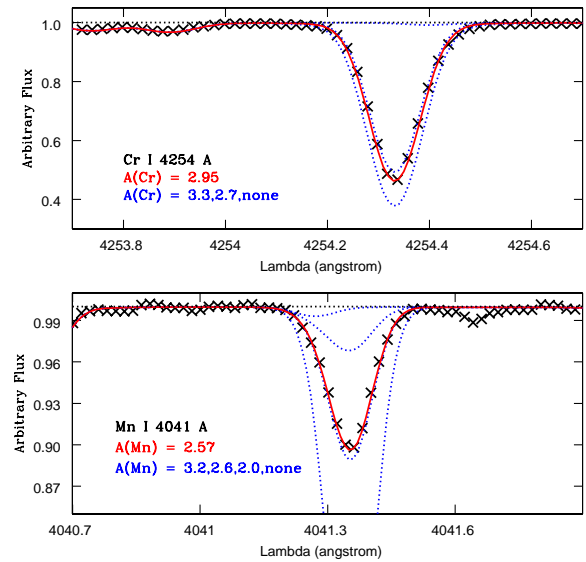}}
\caption{LTE chromium abundance in HD~140283 from \ion{Cr}{I}~4254.33~{\AA} 
line (upper panel) and manganese abundance from \ion{Mn}{I}~4041.35~{\AA} 
line (lower panel). Symbols are the same as in Fig. \ref{Li_fig}.}
\label{Cr_Mn_fig}
\end{figure}

\begin{figure}
\centering
\resizebox{80mm}{!}{\includegraphics[angle=0]{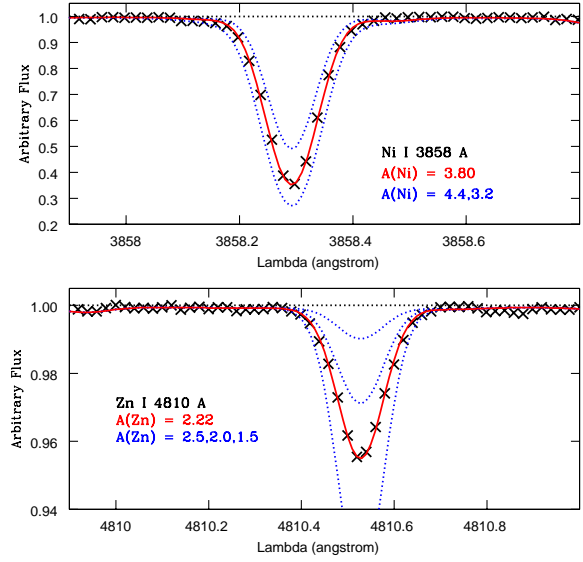}}
\caption{LTE nickel abundance in HD~140283 from \ion{Ni}{I}~3858.29~{\AA} 
line (upper panel) and zinc abundance from \ion{Zn}{I}~4810.53~{\AA} line 
(lower panel). Symbols are the same as in Fig. \ref{Li_fig}.}
\label{Ni_Zn_fig}
\end{figure}

\begin{figure}
\centering
\resizebox{80mm}{!}{\includegraphics[angle=0]{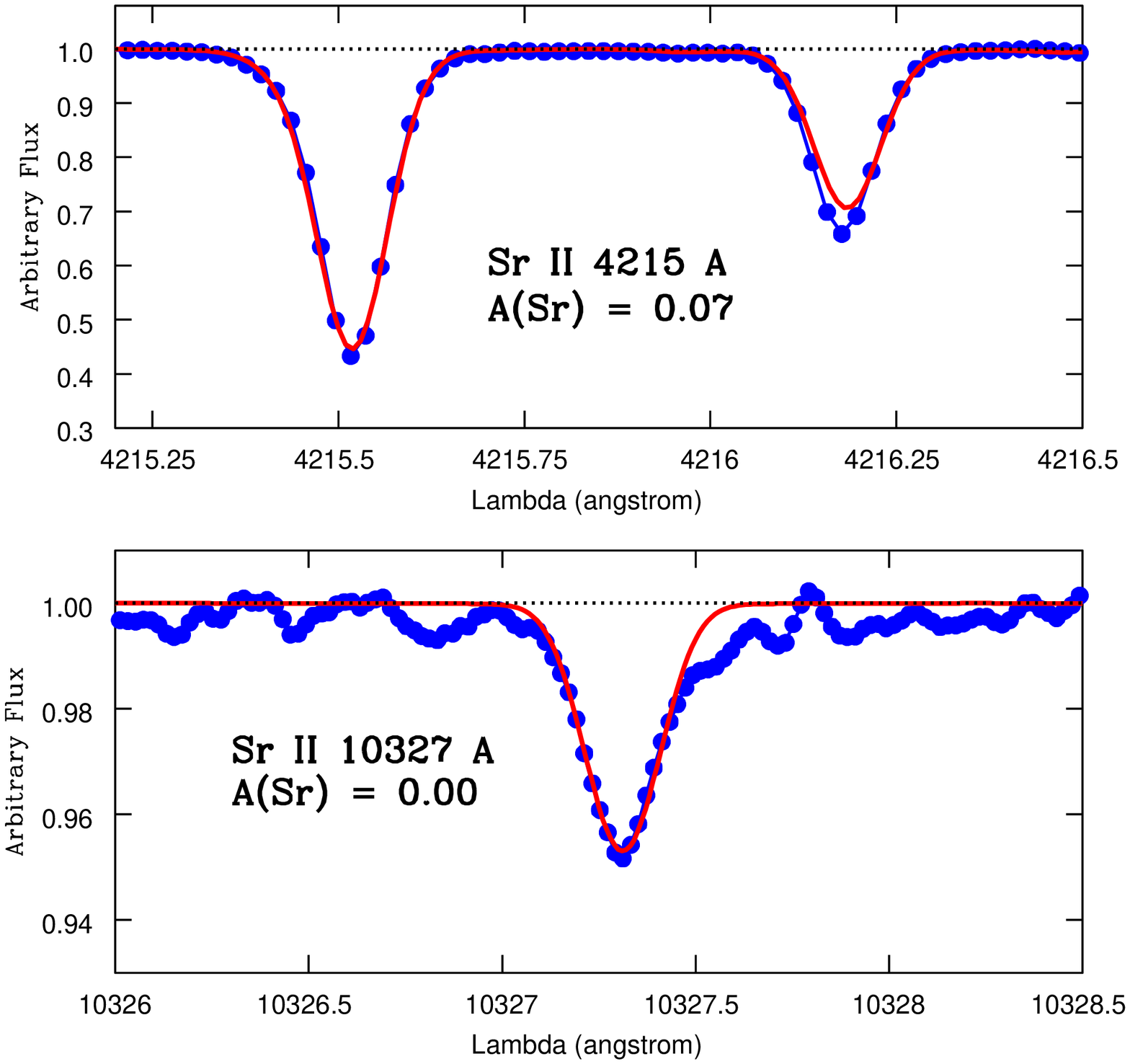}}
\caption{NLTE strontium abundance in HD~140283 from \ion{Sr}{II}~4215.52~{\AA} 
(upper panel) and \ion{Sr}{II}~10327.31~{\AA} (lower panel) lines. 
Symbols are the same as in Fig. \ref{NLTE_O}.}
\label{NLTE_Sr}
\end{figure}

\begin{figure}
\centering
\resizebox{80mm}{!}{\includegraphics[angle=0]{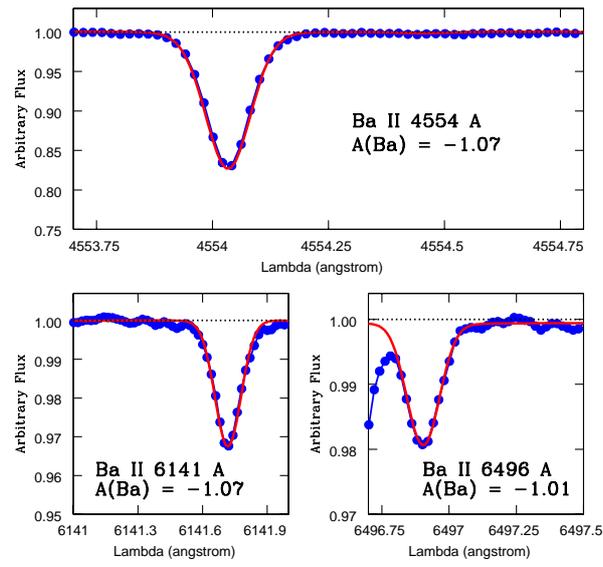}}
\caption{NLTE barium abundance in HD~140283 from \ion{Ba}{II}~4554.03~{\AA} 
(upper panel), \ion{Ba}{II}~6141.70~{\AA} (lower left panel), and 
\ion{Ba}{II}~6496.92~{\AA} (lower right panel) lines. 
Symbols are the same as in Fig. \ref{NLTE_O}.}
\label{NLTE_Ba}
\end{figure}

\end{appendix}

\end{document}